# Is the medium the message?
# Social disclosure channels and firm risk

**Andreas G. F. Hoepner**

Smurfit Graduate Business School, University College Dublin, Belfield, Dublin, Republic of Ireland

Platform on Sustainable Finance, DG FISMA, European Commission, Brussels, Belgium

**Blerita Korca**

University of Bamberg, Faculty of Social Sciences, Economics, and Business Administration

**Frank Schiemann**

University of Bamberg, Faculty of Social Sciences, Economics, and Business Administration

**Fabiola I. Schneider**

(corresponding author)

Smurfit Graduate Business School, University College Dublin, Belfield, Dublin, Republic of Ireland

Platform on Sustainable Finance, DG FISMA, European Commission, Brussels, Belgium

ESMA Sustainability Standing Committee, Paris, France

GHG Protocol Scope 3 Technical Working Group, Washington D.C., United States

*Acknowledgments*:

We are very thankful for comments and inputs from Max Göttsche, Thomas Günther, Markus Heidt, Anne-Kathrin Hinze, Carlota Garcia Manas, Tushar Saini seminar participants at the research seminars at TU Dresden, the University of Essex, the Katholische Universität Eichstätt-Ingolstadt, and conference participants at the 2022 EAA Conference, the 2020 EAERE Conference and the 2019 Irish Academy of Finance Conference in Cork, Ireland. We also would like to thank Donato Calace, Marjella Lecourt-Alma and JP Lecourt of Datamaran for providing us with the sustainability disclosure data. The views expressed in this paper are not necessarily shared by DG FISMA. Andreas Hoepner acknowledges funding from Science Foundation Ireland (Award 19/FIP/AI/7539), Enterprise Ireland (CF-2024-2316-I) and EU Horizon (Grant No. 101182455). All remaining errors are our sole responsibility. Authors are listed alphabetically.

**Abstract**

Investors interpret social disclosures from a risk perspective, yet relevant information can reach them through channels that differ sharply in regulatory enforcement and materiality: SEC filings, sustainability reports, or financial reports. We analyse how social disclosure via each channel relates to idiosyncratic risk. Studying S&P 1,500 constituents, we distinguish between initiated and continued disclosure along the three disclosure channels. We find first-time disclosure of social issues via SEC filings is related to increased idiosyncratic firm risk, highlighting that unexpected information is published. Continuous disclosure of social issues is related to lower idiosyncratic risk for sustainability and financial reports, which is in line with the literature. The SEC filing effect is robust for downside idiosyncratic risk measures, the separation of social disclosure into human capital, product liabilities, and stakeholder engagement, and for propensity score matching. Our findings suggest that the risk impacts of sustainability disclosure depend on the newness of information and disclosure channels.

**Keywords:** Social disclosure, disclosure channels, SEC filings, regulatory disclosure, idiosyncratic risk

**JEL-Classification:** C23; G32; M14; M41

**Highlights**

- We compare social disclosure via SEC filings, sustainability and financial reports
- First-time social disclosure via SEC filings relates to higher idiosyncratic risk
- Continued social disclosure relates to lower idiosyncratic risk for all channels
- Results are robust to downside risk, topic splits and propensity score matching
- Risk effects of disclosure depend on information newness and disclosure channel

# 1. INTRODUCTION

Investors increasingly use sustainability information when assessing firm risk (Amel-Zadeh & Serafeim, 2018; Krueger, Sautner, & Starks, 2020). Yet, social information can reach capital markets through markedly different channels: a highly regulated SEC filing, a voluntary stand-alone sustainability report, or the financial report. These channels differ fundamentally in their degree of regulatory enforcement and their emphasis on financial materiality. If investors interpret social disclosure primarily from a risk perspective, the channel through which the information arrives should shape its effect on firm risk. In this paper, we show that it does: first-time social disclosure via SEC filings is related to increased idiosyncratic risk, whereas continued social disclosure is related to lower idiosyncratic risk.

Corporate sustainability has received increased attention over the last few years (Gillan, Koch, & Starks, 2021; Zhao, Yang, Wang, & Michelson, 2023). Reporting of sustainability-related information plays a central role in this development, as emphasized by initiatives such as GRI (Global Reporting Initiative), IIRC (International Integrated Reporting Council), and the SASB (Sustainability Accounting Standards Board), which were founded to provide guidance on how companies report about their sustainability efforts and performance. Such sustainability disclosures are not mandated in the United States, although the SEC adopted rules on "The Enhancement and Standardization of Climate-Related Disclosures for Investors" in March 2024; the rules were paused amid litigation and the SEC has since proposed their rescission. Under a voluntary disclosure regime, firms retain discretion in the selection of disclosed sustainability issues (Bingler, Kraus, Leippold, & Webersinke, 2022), which can also lead to problems of greenwashing, when firms overstate their sustainability-related practices (Lee & Raschke, 2023). Firms can also decide where to publish sustainability information, for example in their financial reports or in stand-alone sustainability reports. In contrast, the SEC requires the disclosure of financially material information, and since 2005 has required firms to disclose material risk factors, some of which can be sustainability-related in nature.

In our study, we aim to compare the roles of three disclosure channels (i.e., SEC filings, sustainability reports, and financial reports) for sustainability disclosure. Prior research shows that voluntary disclosure can lead to greenwashing behavior, where firms overstate their sustainability-related practices (Lee & Raschke, 2023). Thereby, a stand-alone sustainability report might be more prone to greenwashing than a highly regulated channel, such as SEC disclosure. We are specifically focusing on the relation between social disclosures (in the 2011–2015 period, prior to the politicization of sustainability disclosure) and idiosyncratic risk along the three disclosure channels. We have three reasons for our specific focus. First, we focus on the pre-2016 period to ensure that our estimates capture disclosure channel effects rather than political-risk premia arising from the subsequent polarization of sustainability-related issues. Second, while social disclosure has recently received more attention (Aluchna, Roszkowska-Menkes, Kamiński, & Bosek-Rak, 2022; Lin, Shen, Wang, & Julia Yu, 2024), it still substantially lacks the attention received by environmental and particularly climate disclosures. Third, this study contributes directly to the emerging literature on corporate disclosure channels, asset pricing, and capital market efficiency (e.g., Chen & Xie, 2022; Elshandidy, Fraser, & Hussainey, 2013). While prior accounting and finance research frequently evaluates corporate non-financial disclosures in aggregate, capital market participants do not process information in isolation. Instead, investors evaluate disclosure credibility and financial materiality through channel-dependent filters. By examining how information processing frictions vary between mandatory regulatory filings (e.g., SEC Form 10-K) and voluntary communication channels (e.g., standalone sustainability reports), we show that disclosure location and timing fundamentally alter firm-specific risk profiles. Specifically, our findings demonstrate that cross-channel disclosure strategies interact with market-level information processing, offering novel insights into how non-financial information is priced in capital markets.

In the first step of our analysis, we focus on the general effect of social disclosures on firms' idiosyncratic risks. Thereby, we analyse aggregated social disclosure as well as disclosure on three separate social issues, specifically human capital, product liability, and stakeholder engagement. This means our study is not directly exposed to media criticism of lenient enforcement of SEC climate risk disclosure (Gelles, 2016). Instead, for social issues, such as employee- or product-related issues, there is little doubt surrounding enforceability in the United States courts, which are infamous for punitive damages (Del Rossi & Viscusi, 2010; Garber, 1998). Oftentimes, corporate entities are defendants in cases of punitive damages in the United States courts, and on a large scale plaintiffs receive punitive damage awards regarding serious environmental harms (Del Rossi & Viscusi, 2010). Relatedly, employee lawsuits have been shown to increase stock price crash risk (Zuo, Zhang, Hu, Feng, & Zou, 2022).

In the second step, we contrast firms' social disclosure to the SEC with their voluntary social disclosure integrated into their financial report or reported in a stand-alone sustainability report. We investigate disclosure consistency by simultaneously analysing firm reporting on the same sustainability disclosure topics for the same fiscal year along three disclosure channels. In previous research, the disclosure channel of stand-alone sustainability reports received the most attention. For example, Dhaliwal, Li, Tsang, and Yang (2011) and Dhaliwal, Radhakrishnan, Tsang, and Yang (2012) found that the initiation of sustainability reports themselves reduced analysts' forecast errors, as sustainability reports added complementary information value. We hypothesize that initial SEC disclosure has the opposite effect as it reveals previously unknown risks to the capital market, while we control for the effect of initial disclosure via sustainability reports (in addition to financial reports). According to Brown,

Hillegeist, and Lo  (2009), disclosure does not always decrease uncertainty. On the contrary, disclosure of unexpected accounting information can increase uncertainty. While previous studies have explored individual and alternative disclosure channels for sustainability information (Crowley, Huang, & Lu, 2024; Melloni, Caglio, & Perego, 2017), our study is—to the best of our knowledge—the first empirical investigation of the intrafirm disclosure inconsistencies between simultaneous sustainability disclosure via SEC filings, financial reports, and stand-alone sustainability reports.

By studying S&P 1,500 constituents between 2011 and 2015, we use the Datamaran dataset, which collects non-financial disclosure on a wide range of environmental and social indicators from firms' stand-alone sustainability reports, financial reports, and SEC disclosures. Datamaran applies textual analyses based on machine learning algorithms to systematically analyse a wide range of company reports. Our empirical analysis focuses on the 2011–2015 period, which serves as an ideal 'unconfounded laboratory' to isolate pure channel and timing mechanics prior to post-2016 shifts in institutional ESG mandates and political polarization.

We focus our analysis on initial (i.e., first-time) disclosures of social topics, equivalent to the earlier focus of Dhaliwal et al. (2011) on initial disclosures of social information. While Dhaliwal et al. (2011) consider the initiation of social disclosure in a purely voluntary setting and only focused on disclosure in sustainability reports, in this study, we consider the initiation of social disclosure along three disclosure channels. In particular, the channel of SEC filings introduces a setting in which there is a de-facto mandate to disclose sustainability information if it contains financially material information. We find initial disclosure of social issues, aggregated as well as separately for human capital, product liability, and stakeholder engagement, to the SEC to increase idiosyncratic firm risk. To isolate the underlying mechanism, we further investigate if the risk increase may be linked to the social issue itself instead of the initiated SEC disclosure. Hence, we interact the initial SEC disclosure variable with initial and continued disclosure via another channel (i.e., sustainability report or financial report).

We find that the risk further increases if the same social topic is disclosed simultaneously for the first time via a sustainability report. This result implies that the relevant mechanism is not simply the riskiness of the information disclosed via the SEC and other channels, as observed by Campbell, Chen, Dhaliwal, Lu, and Steele (2014), but more specifically that a combination of initial disclosure via SEC filings and sustainability reports provides a stronger signal of risk. This highlights that unexpected accounting information induces uncertainty (Brown et al., 2009; Rogers, Skinner, & van Buskirk, 2009) and this effect is even stronger when this new information is published via two channels. By contrast, interactions of initial SEC disclosure with continued disclosure via the other channels are negative and significant in some cases, indicating that prior disclosure can mitigate the risk increase. In other words, firms that follow the SEC requirement and initiate disclosure on a social issue, for example on product liabilities, presumably reduce investor trust if they have not previously disclosed the information via other channels. Moreover, if firms initiate disclosure via the SEC and the sustainability channel, they seem to direct additional attention to this trust loss, thus resulting in a further risk increase. Previous literature finds initial sustainability disclosure to be usually risk-reductive (Dhaliwal et al., 2011; Dhaliwal et al., 2012; Plumlee, Brown, Hayes, & Marshall, 2015). However, we find only weak evidence of such a relation for social disclosure via the channels of sustainability reports and financial reports. Consequently, our findings challenge the notion that sustainability disclosure impacts are independent of their disclosure channel and emphasize that investors' perceptions of social information differ depending on the disclosure channel. Specifically, we highlight that when disclosure is characterized by three elements such as (i) disclosure via a highly regulated channel, (ii) financially material information, and (iii) new information which might be surprising, it can increase uncertainty.

Our study makes important contributions to the literature connecting sustainability practices to firm-specific risks (Fafaliou, Giaka, Konstantios, & Polemis, 2022; Kim, Lee, & Kang, 2021; Schiemann & Sakhel, 2019; Zhou & Lei, 2025), including evidence that ESG performance relates to idiosyncratic risk (Sassen, Hinze, & Hardeck, 2016) and shapes corporate risk-taking (He, Ding, Yue, & Liu, 2023). More specifically, we see our first contribution to the literature on the effects of disclosure channels on firms' risk (Blankespoor, deHaan, & Marinovic, 2020), especially by comparing the risk effect of the SEC channel with the sustainability report and financial report channels. This adds to studies focusing on risk disclosure via SEC filings (Campbell et al., 2014; Hope, Hu, & Lu, 2016; Kravet & Muslu, 2013). This further adds to the understanding that investors, who are known to use sustainability-related information (Christophe, Hsieh, & Lee, 2024; Lu, Oh, Kleffner, & Chang, 2021), have a good understanding of the context in which such information is published.

Second, our study also adds to research exploring how sustainability-related regulation impacts company behaviour (Chen, Hung, & Wang, 2018; Christensen, Hail, & Leuz, 2021; Grewal, Riedl, & Serafeim, 2019). In contrast to previous papers, we do not focus on a regulatory change. Instead, we compare a highly regulated disclosure channel (via SEC filings) with channels used for voluntary disclosure (sustainability reports or sustainability information included in financial reports). Thereby, our results show that disclosure via a highly regulated channel can reveal important and otherwise unknown financially material risks to investors.[1]

---

[1] Particularly responsible investors might use such information to increase upside opportunities while constraining downside risk as shown by Gao, Hoepner, Prokopczuk, Rouxelin, and Wuersig (2025).

Third, we also contribute to the literature focusing on information economics (Diamond & Verrecchia, 1991; Grewal et al., 2019; Kothari, Li, & Short, 2009) and risk (Benlemlih & Girerd-Potin, 2017) by providing evidence that social information, which is reported via a regulated channel such as SEC filings, for the first time qualifies as unexpected information, which in turn increases idiosyncratic risk. This adds to the findings from Schiemann and Sakhel (2019), who reveal that the disclosure of climate-related risks can also lead to increased uncertainty.

## 2. SUSTAINABILITY DISCLOSURES ACROSS DISCLOSURE CHANNELS

Interest in sustainability disclosures has increased considerably over the last two decades. Initially, voluntary disclosure standards, such as the GRI, evolved and subsequent research highlighted the positive effects of sustainability disclosures (e.g., Chen & Xie, 2022). For example, Dhaliwal et al. (2011) report a lower cost of equity capital for firms initiating sustainability disclosures, and Dhaliwal et al. (2012) further reveal that the issuance of sustainability reports is related to lower analyst forecast errors. Matsumura, Prakash, and Vera-Muñoz (2014) identify that, on average, firms that voluntarily disclose carbon emissions experience higher firm value compared to non-disclosers.

Although sustainability disclosures have mostly been voluntary in the past, the SEC's materiality principle mandates firms to disclose any information that might influence (potential) shareholders' investment decisions, and sustainability information can qualify as material. To assist in the materiality assessment, the SASB developed a materiality map™ outlining which sustainability topics qualify as material topics for different industries. This has attracted research investigating the relation of material sustainability performance or information to financial performance (Khan, Serafeim, and Yoon, 2016), stock market (Grewal, Hauptmann, and Serafeim, 2021; Serafeim and Yoon, 2022) and firm behaviour (Bochkay, Choi, & Hales, 2022; Göttsche, Griffin, Habermann, Schiemann, & Spandel, 2025).

One can argue that the potential firm value relevance of sustainability-related information in combination with the SEC's requirement to disclose material information can be understood as an implicit mandate to disclose financially material sustainability information via the SEC filings. However, sustainability-related information can be argued to be financially material because potential controversies cause financial risks for companies. This means that financially material sustainability-related information is most likely to be found as part of the risk disclosure, which in itself is criticized for being lengthy and boilerplate (Cazier, McMullin, & Treu, 2021; Kravet & Muslu, 2013). Ibrahim, Hussainey, Nawaz, Ntim, & Elamer (2022) provide a systematic review of this risk disclosure literature.

Firms can report sustainability information via different channels. The three typical channels are stand-alone sustainability reports, financial reports, and SEC disclosures. In stand-alone sustainability reports, firms can choose to follow sustainability reporting standards such as the GRI standards. However, even if firms follow such standards, they still retain significant discretion about what to disclose and how to disclose it. As Hummel and Schlick (2016) show, two directions of the relation between sustainability disclosure and sustainability performance exist simultaneously. Firms with good sustainability performance are likely to report more extensively to highlight their high level of sustainability. At the same time, firms with bad sustainability performance report more to obfuscate their subpar performance. Consistent with such opportunism, Farag, Meng, & Mallin (2015) find, based on a social, environmental, and ethics disclosure index for Chinese listed companies, that better financial performance is associated with lower social disclosure. Therefore, readers of voluntarily published sustainability reports might assume greenwashing through overly optimistic representations of firms' environmental and social performance. For example, Cho, Lee, and Pfeiffer (2013) analyse how positive and negative sustainability information is associated with information asymmetry and report that information asymmetry is only significantly smaller for firms with more negative sustainability information. They interpret this to mean that negative information is perceived to be more credible, while positive information seems to be largely ignored. Similarly, Schiemann and Sakhel (2019) ascertain that increased reporting of climate-related physical risks is related to lower information asymmetry, while increased reporting about climate-related opportunities is not. In a similar vein, Schiemann & Tietmeyer (2022) show that ESG disclosure mitigates the increase in analyst forecast errors associated with ESG controversies, with the social pillar being the most important driver of this relation. Voluntary disclosure schemes provide more opportunities for over-optimistic disclosures (Bingler et al., 2022), while highly regulated disclosure channels, such as SEC filings, might mitigate such tendencies. Furthermore, sustainability reports often address a wider audience and are thus not focused on shareholders specifically. For example, studies focusing on sustainability-related reputational risks (Christophe et al., 2024; Fafaliou et al., 2022) at least implicitly acknowledge that customers can be impacted by such information. Similarly, activist groups can put pressure on firms and threaten their legitimacy if sustainability-related problems become apparent (Lee and Raschke, 2023). This is also acknowledged by sustainability reporting standards: The GRI standard specifically focuses on stakeholder orientation in its guiding principles and defines materiality as information that is relevant to any stakeholder of the firm. Therefore, shareholders might find it hard to understand the financial implications of non-financial information provided via a sustainability report.

On the contrary, the SEC disclosures demand the disclosure of financially material information. Put simply, sustainability-related information should be disclosed via SEC filings if it is linked to potential financial consequences for the disclosing company. The SASB standards, published from 2013 to 2016, indicate financially material sustainability topics for different sectors. Relying on SASB's materiality map, Khan et al. (2016) show that firms with good ratings on their material sustainability issues perform significantly better than firms with poor ratings on such issues, while this significant difference is not found for immaterial sustainability issues. In SEC filings, firms are likely to report sustainability information only if it is financially material. This is often the case for sustainability-related risks (e.g., physical risks from climate change, regulatory risks due to carbon prices or carbon taxes).[2] Berkman et al. (2024) and Matsumura et al. (2024) analyse climate-related information published via SEC's 10-K filings and confirm significant relations between the disclosure of such information and firm value, as well as firm risk. Relatedly, Ignatov (2023) finds that the social and governance tone of 10-K reports is significantly related to stock returns around filing dates.

Sustainability information can also be included in the financial report. Typically, a company would choose to report more extensively about ESG issues in the management discussion, where the firm can provide not only quantitative information but also narratives about current projects or plans to improve the sustainability performance of a firm. Some international regulatory efforts have placed more emphasis on this disclosure channel. For example, South Africa introduced a mandate for integrated reporting in 2010 (Barth, Cahan, Chen, & Venter, 2017) and the EU's Nonfinancial Disclosure Directive (2014/95/EU) mandates the disclosure of certain sustainability-related information in the financial report or in a separate sustainability report.[3] However, in the US setting, including sustainability information in the financial report is a voluntary choice, except for those issues, which are clearly financially material. Naturally, ESG information disclosed via financial reports is aimed at capital markets. However, firms can exercise discretion regarding what to disclose and firms can also choose how to disclose related information, similar to voluntary stand-alone sustainability reports.

---

[2] For example, in 2010 the SEC published the "Commission Guidance Regarding Disclosure Related to Climate Change" (Release Nos. 33-9106; 34-61469; FR-82). Current developments are summarized at the SEC's website: https://www.sec.gov/sec-response-climate-and-esg-risks-and-opportunities.

[3] The Corporate Sustainability Disclosure Directive (CSRD) in the EU is a further development of the NFRD and proposes, amongst other things, that sustainability disclosures become part of the management report.

Based on the aforementioned explanations, it seems reasonable that firms make decisions about what sustainability issues they address and which pieces of information they disclose, at least in part based on the disclosure channel (e.g., more narrative, anecdotal information via sustainability reports, more risk-related information via SEC filings). At the same time, recipients of the social disclosure process evaluate the information differently, depending on which disclosure channel was used by the disclosing firm. For example, information provided via SEC filings might be seen as more credible and the associated risks might be perceived as more tangible. Our study aims to analyse whether disclosures via SEC filings are, indeed, perceived differently from other disclosure channels, such as sustainability reports or financial reports.

Christensen, Floyd, Liu, and Maffett (2017) demonstrate that disclosures of mine-safety records via the SEC are perceived as important information by capital market participants and have incremental real effects, although such records are publicly available via other channels as well. This example illustrates that disclosures via the SEC differ in how capital market participants acknowledge and process the information. In this regard, Blankespoor et al. (2020) emphasize that little is known about the effects of disclosure via SEC vs. other channels, and further research is warranted. Recent evidence also indicates that disclosure channels are interdependent: Lin, Wang, Wu, & Wu (2026) show that mandating standardized disclosure through one channel can inhibit the information supplied through another (in their setting, by auditors) and weaken the market response to that information. While Christensen et al. (2017) focus on the real effects of mandatory disclosure via the SEC, we are interested in the capital market effects stemming from mandatory disclosure via the SEC. Therefore, in contrast to Christensen et al. (2017), we explore the impact of first-time SEC disclosure, in comparison to other channels, following the call for research on this issue by Blankespoor et al. (2020). To the best of our knowledge, this is the first study that provides empirical evidence on the relationship between first-time SEC disclosure and idiosyncratic risk.

## 3. HYPOTHESIS DEVELOPMENT

Different disclosure channels have differing effects on the capital market (see Blankespoor et al., 2020; Christensen et al., 2017; Elshandidy et al., 2013; Guidry & Patten, 2010). Social disclosure via SEC filings provides a unique setting in this regard. Accordingly, we explore first-time SEC disclosure and argue that it has significant effects on capital market participants. As we argue below in more detail, this is likely to occur due to first-time SEC disclosure being i) highly regulated (especially compared to providing sustainability information in stand-alone reports or financial reports), ii) financially material information, and iii) new information which can be surprising for capital market participants.

Investors demand high-quality, objective and standardized disclosure (Guidry & Patten, 2010; Ilhan, Krueger, Sautner, & Starks, 2023). While voluntary sustainability reporting is considered incomplete and lacking objectivity (Unerman, Bebbington, & O'Dwyer, 2007), on the contrary, mandatory disclosure is aimed at enhancing information quality. Therefore, to fulfil their information needs, investors pay more attention to disclosure in a regulated setting (Schiemann and Sakhel, 2019). For SEC filings, firms have to follow the strict format provided by the SEC. Thereby, sustainability information can become relevant, for example, as part of the risk disclosure within the 10-K statement. In comparison to sustainability disclosure via a stand-alone report or as part of the financial report, the structure of SEC filings allows for the lowest degree of discretion among the three aforementioned channels. Accordingly, social disclosure provided via SEC filings is most likely financially relevant to (potential) shareholders (Christensen et al., 2017).

In addition to the highly regulated nature of SEC filings, the focus on financial materiality plays an important role in how the disclosed social information is perceived by investors. In this regard, the guidance provided by the materiality map™ of SASB is specifically focused on SEC filings and financial materiality. Therefore, firms have clear indications of what information they should report. Campbell et al. (2014) found that disclosure via SEC indeed contains financially material information such as meaningfully reflecting the risks a firm faces. Similarly, Beatty, Cheng, and Zhang (2019) found a significant positive relation between unexpected risk factor disclosures and investor reactions in the equity markets around 10-K filings. In this regard, previous literature shows that financially material information such as climate-related risk disclosure published via SEC filings has a significant relation with firm value (Berkman et al., 2024; Matsumura et al., 2024). Similarly, social disclosure is typically interpreted from a risk perspective by capital markets and firms are likely aware of this (Benlemlih, Shaukat, Qiu, & Trojanowski, 2018). The degree of materiality definition can also be associated with the level of specificity provided in the SEC filings. According to Hope et al. (2016), the specificity of firms' risk disclosure in 10-K filings is related to a stronger market reaction. Therefore, in this study, we also argue that SEC filings, which follow a financial materiality approach, will attract market reaction due to the specificity of information which is of interest to investors.

Discussions about the relation between sustainability disclosures and the information environment typically contrast two scenarios (e.g., Cho et al., 2013; Dhaliwal et al., 2011; Dhaliwal et al., 2012): (1) Sustainability information is useful for the capital market and, hence, significantly improves the information environment vs. (2) sustainability information is not useful and, hence, no significant relation between information environment and sustainability disclosure can be found. These contrasting views imply a third scenario: sustainability information impairs the information environment. Empirically, the relation between CSR and idiosyncratic volatility is indeed not uniformly negative (Becchetti, Ciciretti, & Hasan, 2015). Such a scenario requires that the information provided is unexpected (Brown et al., 2009; Kothari et al., 2009). Considering that SEC disclosure is more regulated than disclosure via the other channels, and contains financially material information, which, as argued above, attracts investors' interest (Hope et al., 2016), especially when happening for the first time, it can come as a surprise to investors. Previous research has shown that surprising or unexpected accounting information has the potential to induce uncertainty (Brown et al., 2009; Rogers et al., 2009; Schiemann & Sakhel, 2019). Schiemann and Sakhel (2019) show that, for climate-related disclosures about physical risk, in some cases disclosures about higher risk exposures are related to higher bid-ask spreads, indicating that such information increases uncertainty. Subsequently, we argue that initial social disclosure via SEC disclosures will also qualify as this type of information.

Our arguments in this paper emphasize that first-time SEC disclosure is characterized by three elements which co-exist (highly regulated disclosure, financially material information, and new information which might be perceived as surprising). Taking into account the arguments developed above regarding each of those elements characterizing SEC disclosure, the following hypothesis is proposed:

*$H_{1a}$: First-time social disclosure via SEC filings is related to increased idiosyncratic risk*

The theoretical mechanism underpinning $H_{1a}$ rests on the distinction between voluntary narrative framing and legally binding material risk recognition. Under U.S. federal securities laws, incorporating a social issue into SEC mandatory filings (*e.g.*, Item 1A Risk Factors) imposes strict legal liability under Rule 10b-5. Consequently, a firm's first-time disclosure of a social topic in an SEC filing signals to market participants that management now considers this issue to be financially material, transitioning the topic from a discretionary corporate responsibility narrative to a potential latent liability. Rather than simply resolving information asymmetry, this initial recognition triggers investor uncertainty regarding the magnitude of latent costs, potential litigation exposures, and required capital expenditures. This fundamental re-pricing of firm-specific risk and cash-flow uncertainty ultimately manifests as a short-term spike in idiosyncratic volatility.

In contrast to our study, previous research has empirically assessed how the initiation of voluntary sustainability reports can impact risk. While we argue that SEC filings, which are more regulated than sustainability reports and which focus on financially material information, can increase risk, studies in the past found that disclosure via the channel of sustainability reports can be risk-reductive (Dhaliwal et al., 2011; Dhaliwal et al., 2012). Therefore, those studies account for voluntary reporting, which is often associated with a "cherry-picking" approach (Bingler et al., 2022). This means that companies can choose to disclose the most favourable information, which is most likely to reduce the uncertainty of investors. Similarly, Matsumura et al. (2024) highlight that, in a voluntary setting, firms can evaluate the benefits versus costs of disclosing risk-related information which then impacts their reporting decisions. In other words, regulatory enforcement, for example, to facilitate balanced reporting of positive and negative news, is limited in voluntary disclosure settings. While this is especially true for sustainability reports, companies might face more scrutiny when including sustainability information in their annual reports. However, even in these cases, the disclosure of sustainability information is voluntary for the most part. For this reason, we argue that SEC disclosure represents a channel with a highly regulated nature and a strong emphasis on financial materiality of the information provided. To formalize these theoretical mechanisms, Figure 1 outlines our conceptual framework along two primary dimensions: Regulatory Enforcement (high vs. low) and Financial Materiality (perceived vs. binding). Quadrant I (High Enforcement / Binding Materiality) represents SEC regulatory filings, where first-time disclosures ($H_{1a}$) act as binding signals of latent liability, temporarily increasing idiosyncratic risk. Quadrant IV (Low Enforcement / Perceived Materiality) encompasses voluntary sustainability reports, where continuous disclosure ($H_{2b}$) serves an information-resolution role, smoothing uncertainty and decreasing firm-specific risk over time. Cross-Quadrant Interactions ($H_{1c}$ and $H_{2c}$) illustrate how prior voluntary signaling in Quadrant IV mitigates the initial volatility shock when management eventually transitions a social topic into Quadrant I regulatory filings.

[Insert Figure 1 about here.]

While we argue above that first-time social disclosure via SEC filing is characterized as being a highly regulated disclosure channel, providing financially material information, which at the first time can have a surprising effect on (potential) shareholders, other disclosure channels such as financial reports and sustainability reports are not characterized simultaneously by all those elements. Figure 1 emphasizes that only SEC filings are characterized by a high level of materiality definition and regulatory enforcement, while the other two channels are not. While both financial reports and stand-alone sustainability reports can contain unexpected or surprising news when first-time social disclosure occurs, they lack the other two elements (being highly regulated and containing financially material information). Therefore, the following hypotheses are proposed:

*$H_{1b}$: First-time social disclosure via stand-alone sustainability reports has no significant effect on idiosyncratic risk*

*$H_{1c}$: First-time social disclosure via financial reports has no significant effect on idiosyncratic risk*

Furthermore, we account for the "newness" element in SEC disclosure, which is an addition to previous literature analysing how the existence of prior or continuous sustainability information helps to mitigate risk (Benlemlih et al., 2018; Blacconiere & Patten, 1994; Freedman & Patten, 2004). Theoretically, Albuquerque, Koskinen, & Zhang (2019) model CSR as a product differentiation strategy that reduces firm risk. Benlemlih et al. (2018) studied the link between firms' social disclosure and idiosyncratic risk. The authors consider continuous social disclosure and how that helps to mitigate idiosyncratic risk. Their findings suggest that extensive and objective social disclosure is negatively associated with a firm's total and idiosyncratic risk. Similarly, Blacconiere & Patten (1994) studied the market reaction after the chemical leak in Bhopal, India, and found that prior environmental disclosure in 10-K reports helped US chemical firms mitigate the extent of the negative market reaction. Freedman & Patten (2004), studying the impact of the Toxics Release Inventory (serving as quasi-regulation), found that companies with more extensive environmental disclosure suffered less negative market reactions than those with less disclosure. Therefore, considering the arguments developed above, the following hypotheses are proposed in this study:

$H_{2a}$: *Continuous social disclosure via SEC filings is related to decreased idiosyncratic risk*

$H_{2b}$: *Continuous social disclosure via stand-alone sustainability reports is related to decreased idiosyncratic risk*

$H_{2c}$: *Continuous social disclosure via financial reports is related to decreased idiosyncratic risk*

## 4. RESEARCH DESIGN

### 4.1 Data

Our study relies on data from Datamaran, which captures firm-specific sustainability disclosure from a large sample of US companies and distinguishes between the disclosure channels of sustainability reports, financial reports, and SEC filings. For this purpose, Datamaran automatically scans the documents related to each disclosure channel and applies machine learning algorithms for textual analysis. Given the discretion that researchers experience in designing machine learning algorithms per se (Arimond et al., 2020; Gu, Kelly, & Xiu, 2020) and text extraction in particular (Chen, Wu, & Yang, 2019; Thng, 2019), using Datamaran similar to previous literature (Aksoy, Buoye, Fors, Keiningham, & Rosengren, 2022; Hoepner, Majoch, & Zhou, 2021) instead of developing our own machine learning algorithms prevents us from exposure to look-ahead bias. In other words, if we designed our own fact extraction algorithm (e.g., word2vec), we would be in full control of the data generation process ourselves. Consequently, we could use our knowledge to *look ahead and game* our machine-learning specifications in a manner that would eventually lead to our desired result. Abstaining from the development of a personal suite of machine learning algorithms is furthermore in line with Berkman, Jona, and Soderstrom (2024), who used an externally extracted data sample, and Matsumura, Prakash, and Vera-Muñoz (2024), who hand-collected their data. Datamaran identified 109 different ESG items and for every firm also identified which of these ESG items are mentioned via each disclosure channel.

While the Datamaran dataset provides a detailed picture of firms' sustainability disclosure, this leads to very fragmented results when focusing on individual sustainability items. Therefore, we focus on the social pillar, specifically on three main social topics: "Human capital", "Product liabilities", and "Stakeholder engagement". We identify Datamaran's corresponding social items and aggregate these items into each of the three social topics. We coded 20 social items for "Human capital"[4], 11 social items for "Product liabilities"[5], and 15 social items for "Stakeholder engagement"[6]. Furthermore, we create binary variables to assess whether the respective social issue is addressed in firm disclosure via SEC, financial report, and/or sustainability report. Consequently, our binary disclosure score for each of the three social issues and each of the three reporting channels is 0 for "no reporting" (i.e., if all Datamaran items for the respective social topic reveal no disclosure) and 1 for "reporting" (i.e., if at least one Datamaran item for the social topic reveals disclosure). In conclusion, for each of the three reporting channels and each of the three social issues, we have one binary variable capturing whether or not a firm disclosed in any of the specific years during our sample period.

Our focus is on first-time disclosures via SEC filings. Building on the binary disclosure variables for each disclosure channel, we further create a variable capturing first-time disclosure. In the first step, we take a broad focus on any social disclosure and create the variables SEC_First, SUS_First, and FIN_First, which equal one if a company disclosed information on any social issue via SEC, sustainability report, and financial report, respectively, in the current year, but not in any of the previous years in our sample. Furthermore, we create topic- and channel-specific first-time disclosure variables, resulting in nine variables (i.e., SEC_Human_First, SEC_Product_First, SEC_Stakeholder_First, SUS_Human_First, SUS_Product_First, SUS_Stakeholder_First, FIN_Human_First, FIN_Product_First, FIN_Stakeholder_First). Variables equal one if a company reports about the respective social issue via the respective channel in the current year, but not in any previous year, and are equal to zero otherwise (i.e., for non-reporters as well as for repeated reporters). In the final step, we aim to include control variables in our regression models to distinguish non-reporters from firms with repeated disclosures. Therefore, we also create dummy variables to capture continued reporting. These variables follow the same structure as "first disclosure" variables (i.e., SEC_Cont, SUS_Cont, FIN_Cont; and SEC_Human_Cont,… FIN_Stakeholder_Cont). Note that continued disclosure variables equal one if a firm reports via the respective channel (and on the respective social issue) in the current year and did so in at least one previous year in our sample.

---

4 The items for "Human capital" are Children's rights, Workforce diversity & inclusion, Employee benefits, Employee development, Employee engagement, Employee satisfaction, Employee volunteering, Employment general, Fair remuneration, Forced labour, Human rights, Human trafficking, Labor rights, Sexual exploitation, Skilled workforce, Social inclusion, Supply chain engagement, Supply chain management, Unionization, and Workforce changes.

5 The items for "Product liabilities" are Consumer rights, Customer satisfaction, Customer privacy & security, Grievance mechanisms, Harmful substances, Information security, Lifecycle Management - Products & Services, Lifecycle - Products & Services, Product & service safety, Product stewardship, and Product take-back.

6 The items for"Stakeholder engagement" are Relief & disaster aid, Community engagement, Community impact, Community support & development, Digital inclusion, Financial inclusion, Financial literacy, Local economy, Local infrastructure investment, Market access, Net neutrality, Nutrition, Occupational health & safety, Philanthropy, and Shareholder activism.

### 4.2 Determinants of Firm Risk

We aim to explain firms' idiosyncratic risk with the following regression models (no subscripts provided to facilitate readability):

All channels:

$$\begin{aligned} Id_risk = {} & \alpha_0 + \alpha_1 SEC_First + \alpha_2 SUS_First + \alpha_3 FIN_First + \alpha_4 SEC_Cont \\ & + \alpha_5 SUS_Cont + \alpha_6 FIN_Cont + \alpha_7 Size + \alpha_8 ROA + \alpha_9 SDROA \\ & + \alpha_{10} Loss + \alpha_{11} MTB + \alpha_{12} AnCov + \alpha_{13} ESGScore + \alpha_{14} SP500 \\ & + \alpha_{15} FinTran + \alpha_{16} FinTraInt + Industry\text{-}controls + \varepsilon_1 \qquad (1) \end{aligned}$$

Specific channels:

$$\begin{aligned} Id_risk = {} & \beta_0 + \beta_1 [Channel]_First + \beta_2 [Channel]_Cont + \beta_3 Size + \beta_4 ROA \\ & + \beta_5 SDROA + \beta_6 Loss + \beta_7 MTB + \beta_8 AnCov + \beta_9 ESGScore + \beta_{10} SP500 \\ & + \beta_{11} FinTran + \beta_{12} FinTraInt + Industry\text{-}controls + \varepsilon_2 \qquad (2) \end{aligned}$$

We calculate firm and year-clustered standard errors. We apply idiosyncratic risk as the dependent variable. The calculation follows Fu (2009). For each firm, we regress the daily returns on their respective Fama and French (1993; 1996) three factors (retrieved from Kenneth R. French's website[7]). The idiosyncratic risk is the standard deviation of the error term of that regression across one year.

Our variables of interest are the first-time disclosure variables *[Channel]_First*, where *Channel* indicates the respective disclosure channel of SEC filings (SEC), sustainability reports (SUS), and financial reports (FIN).

We apply further control variables addressing firm size (*Size*), profitability (*ROA*), the standard deviation of profitability (*SDROA*), loss-dummy (*Loss*), market-to-book ratio (*MTB*), analysts following the firm (*AnCov*), overall ESG score (*ESGScore*), S&P500-dummy (*SP500*), and financial transparency (*FinTran* & *FinTraInt*). All variable definitions and calculations are explained in detail in Table 1. All continuous variables are winsorized at the top and bottom 1 per cent.

[Insert Table 1 about here.]

We further analyse disclosure for each of the three social topics separately based on regression model (1), in which we substitute the *[Channel]_First* and *[Channel]_Cont* variables with the [*Channel]_[Topic]_First* and *[Channel]_[Topic]_Cont* variables. Furthermore, we use interaction variables to test whether first-time disclosure via SEC in combination with first-time disclosure via sustainability or financial reports is related to stronger or weaker increases in idiosyncratic risk. In these models, we also control for similar interactions of continued disclosures. Accordingly, the models tested take the following form:

$$
\begin{aligned}
Id_risk = {} & \gamma_0 + \gamma_1 SEC_[Topic]_First + \gamma_2 SUS_[Topic]_First + \gamma_3 FIN_[Topic]First \\
& + \gamma_4 SEC_[Topic]_First \times SUS_[Topic]_First \\
& + \gamma_5 SEC_[Topic]_First \times FIN_[Topic]_First \\
& + \gamma_6 SEC_[Topic]_Cont + \gamma_7 SUS_[Topic]_Cont + \gamma_8 FIN_[Topic]_Cont \\
& + \gamma_9 SEC_[Topic]_Cont \times SUS_[Topic]_Cont \\
& + \gamma_{10} SEC_[Topic]_Cont \times FIN_[Topic]_Cont \\
& + \gamma_{11} Size + \gamma_{12} ROA + \gamma_{13} SDROA + \gamma_{14} Loss + \gamma_{15} MTB + \gamma_{16} AnCov \\
& + \gamma_{17} ESGScore + \gamma_{18} SP500 + \gamma_{19} FinTran + \gamma_{20} FinTraInt \\
& + Industry\text{-}controls + \varepsilon_3 \qquad (3)
\end{aligned}
$$

[7] http://mba.tuck.dartmouth.edu/pages/faculty/ken.french/data_library.html. We thank Kenneth French for making the data available.

**4.3 Sample and Data Sources**

We analyze a sample of 755 US companies ranging from 2011 to 2015 with a total of 2,982 firm-year observations. Our sample period spans 2011 through 2015, providing two distinct methodological advantages for isolating the capital market effects of corporate disclosure channels. First, this window precedes the formal codification and widespread institutional adoption of the Sustainability Accounting Standards Board (SASB) sector standards post-2015, capturing an era when management exercised genuine discretion over both topic selection and channel choice. Second, as highlighted in recent literature (e.g., Moretti, Terzani, & De Novellis, 2026; Schroeder & Ormazabal, 2024), post-2016 corporate ESG disclosures became increasingly confounded by macro-political polarization and systemic ESG risk premia. By restricting our empirical setting to 2011–2015, we ensure that our estimates capture pure information-processing frictions and channel credibility, unconfounded by exogenous political regime shifts or standardized reporting mandates. Empirical studies treat pre-2016 and post-2016 as distinct disclosure regimes (Moretti et al., 2026). In such a regime, relations between disclosure initiation and idiosyncratic risk would conflate the information-processing effects we seek to identify with the consequential polarization of sustainability-related issues.[8] Starting with the Datamaran dataset, we add fundamental firm information from Compustat, capital-market information from CRSP, data on analyst following from IBES, and the MSCI ESG Intangible Value Assessment (IVA) score. The final dataset (2,982 firm-year observations) contains all firm-year observations for which all relevant information is available.

## 5. RESULTS

### 5.1 Descriptive Statistics

Table 2 reports descriptive statistics for the dependent variable, the variables of interest, as well as control variables. Panel A focuses on the dependent variable and control variables, and panel B reports the frequencies of first-time disclosure along the three disclosure channels and the social topics. We report an average idiosyncratic risk of 0.014. Firm-year observations, on average, have a return on assets (ROA) of 9.5%, and 6.7% of all observations belong to loss years. Furthermore, 55.0% of all observations are from firms contained in the S&P 500 in the respective year. Panel B shows that the frequency of first-time disclosure is highest for the SEC channel (6.5%), followed by SUS (5.9%) and FIN (3.0%). Therefore, topic-specific first-time disclosures are generally highest for SUS, with product liabilities and stakeholder engagement showing 5.1% of first-time disclosures and human capital at 5.0%. While the descriptive statistics reveal that first-time disclosures occur for every channel and across all three social topics, it does not seem to be consistent across the disclosure channels.

[8] Note, our Datamaran data ranges from 2010 to 2015. Therefore, we can start to measure first-time disclosure variables in 2011 (if in 2010 no data was reported, but there is data reported in 2011). However, as no data prior to 2010 is available (see section 4.1) from Datamaran, we have to use 2010 as the base year to measure first-time disclosure. Extending the sample into more recent years would, in any case, require confronting a further regime break: during the COVID-19 pandemic, workforce-related developments (e.g., remote work, layoffs, workplace safety) became an acute source of idiosyncratic return variation, which would confound the relation between our human capital measure and idiosyncratic risk.

[Insert Table 2 about here.]

### 5.2 Regression Models

Table 3 reports the results of regression models (1) and (2), which regress idiosyncratic risk on first-time and continued disclosure for (each of) the three disclosure channels. For first-time disclosures, the table reports significant results only for the SEC channel, meaning support for $H_{1a}$, but not for $H_{1b}$ and $H_{1c}$. The coefficient for SEC_First is positive and significant in column (1) for the model including all disclosure channels (coeff.: 0.0531; $p < 0.01$) and in column (2) for the model focusing on the SEC channel (0.0542; $p < 0.01$). For continued disclosures, we find significant results with negative coefficients for $H_{2b}$ and $H_{2c}$, meaning the channels sustainability report (-0.0338; $p < 0.1$ in column (1), and -0.0422; $p < 0.05$ in column (3)) and financial report (-0.0542; $p < 0.1$ in column (1), and -0.0592; $p < 0.05$ in column (4)). However, $H_{2a}$ is not supported as continued disclosure via the SEC channel shows a negative but insignificant coefficient (-0.0387; $p > 0.1$ in column (1), and -0.0318; $p > 0.1$ in column (2)).

The results support our expectation formulated in $H_{1a}$ that first-time disclosures via SEC filings are risk-increasing. However, first-time disclosures via the other two channels are not significantly related to firm risk (i.e., no support for $H_{1b}$ and $H_{1c}$). These results contrast with prior literature, suggesting that the first-time issuance of a sustainability report reduces uncertainty (Dhaliwal et al., 2011). Relatedly, He, Qin, Liu, & Wu (2022) find that the initiation of ESG information disclosure via CSR reports is related to lower idiosyncratic risk in China. Instead, for continued disclosure we find risk-decreasing effects for social disclosures via both sustainability reports ($H_{2b}$) and financial reports ($H_{2c}$). This is in line with previous literature, which finds risk-decreasing effects for continued disclosure (Benlemlih et al., 2018; Blacconiere & Patten, 1994; Freedman & Patten, 2004). It has to be noted, however, that our focus on the disclosure of more detailed sustainability topics (i.e., social disclosures) is somewhat different from the general issuance of a sustainability report, where firms have discretion over the specific sustainability topics they disclose. Accordingly, our results also indicate that the relation between disclosure about specific sustainability topics and uncertainty can differ from the relation between the much more aggregated decision to issue a stand-alone sustainability report and uncertainty.

[Insert Table 3 about here.]

[Insert Table 4 about here.]

Table 4 reports the results for social disclosure along three separate social issues and with (downside) idiosyncratic risk as a dependent variable. We report positive and significant ($p < 0.01$) coefficients for SEC_[Topic]_First in all six models, which provides further support for $H_{1a}$ and indicates that first-time social disclosure via the SEC is related to higher (downside) idiosyncratic risk. At the same time, for SUS_[Topic]_First, we find no significant coefficients, and for FIN_[Topic]_First, four out of six models show a negative and significant ($p < 0.1$) coefficient. Consequently, we discover weak evidence at best for a relation between first-time disclosures via financial ($H_{1c}$) or sustainability ($H_{1b}$) reports and this relation, if significant, is negative. These results emphasize the role of SEC disclosures as the only channel with risk-increasing and consistent results. As highlighted above, SEC disclosure not only represents a highly regulated disclosure channel, but also focuses on financially material information. Therefore, SEC filings are of high interest to investors, and first-time social disclosure can be perceived as surprising information, which in turn can increase uncertainty.

For continued disclosure, we find positive and significant coefficients for the SEC channel, SEC_Human_Cont, for the model estimating idiosyncratic risk (0.0568; $p < 0.1$) and for the model estimating negative idiosyncratic risk (0.0822; $p < 0.01$). For sustainability reports, we find no significant results for continued disclosures, and for financial reports, we find negative and significant coefficients for all models estimating idiosyncratic risk. Therefore, support for $H_{2a}$ through $H_{2c}$ does not seem robust over a range of different models and social issues, whereby for $H_{2c}$ (financial reports) we report the most consistent results across the three social issues.

Overall, the results strongly support our expectations that first-time SEC disclosure increases idiosyncratic risk. This is especially noteworthy because the SEC_[Topic]_First variables are the only disclosure variables with consistent and highly significant results. Variables for the other disclosure channels and variables covering continued disclosures return insignificant or varying results across the three social issues.

[Insert Table 5 about here.]

Table 5 reports results for analyses following regression model (3), which takes into account interdependencies of the disclosure channels by including interaction variables. As our focus is on the initiation of the SEC disclosure channel, we interact SEC_[Topic]_First with the first-time and continued disclosure variables for sustainability and financial reports. First, we identify that the directions and significances of SEC_[Topic]_First remain consistent, as they are positive and highly significant ($p < 0.01$) for all three social issues. Second, for the interaction terms, we find consistent and significant results only for the interaction SEC_[Topic]_First and SUS_[Topic]_First: If firms report about each of the three social topics for the first time via both sustainability reports and SEC filings, we see a highly significant ($p < 0.01$) increase in idiosyncratic risk. In other words, there is an additional risk-increasing effect if a firm does not merely report social information for the first time via the SEC but also via its sustainability report. Considering the argument that stand-alone sustainability reports provide more details on sustainability topics (Dhaliwal et al., 2011), our results can be interpreted to mean that these increased details do not necessarily mitigate risk but instead enhance the "surprise" effect on investors. This is also in line with Kravet and Muslu (2013), who report a "divergent effect" for risk disclosure, meaning that more disclosure increases uncertainty at the stock market.

We also find that first-time SEC disclosure in parallel to continued disclosure via other channels can have a risk-decreasing effect in some cases: SEC_[Topic]_First x SUS_[Topic]_Cont is negative and significant for stakeholder engagement (-0.0653, $p < 0.1$), and SEC_[Topic]_First x FIN_[Topic]_Cont is negative and significant for product liabilities (-0.2063, $p < 0.1$) and stakeholder engagement (-0.2483, $p < 0.05$). This indicates that, in some cases, firms can counter the negative effect of surprising investors with new social information reported in their SEC filings by addressing the issues beforehand in their annual financial or sustainability reports. Thus, in cases where disclosure is not characterized by the three elements (highly regulated disclosure, financially material information, and newness of information), risk can be mitigated. In this case, the newness of information can be countered by prior sustainability-related disclosure to mitigate the effect of surprising investors. This finding is in line with prior literature that the existence of prior sustainability-related disclosure helps to mitigate risk (Benlemlih et al., 2018; Blacconiere & Patten, 1994; Freedman & Patten, 2004).

The results for all regression models explaining idiosyncratic risk strongly support $H_{1a}$, and suggest that SEC disclosures, in general, are meaningful for the capital markets and reveal unexpected news. Firms can voluntarily decide whether to disclose social information via their sustainability or financial report and what specific information items to include in such a report. This might lead to the disclosure of financially immaterial information, for example, in a narrative form with little to no relevance for capital markets. However, for SEC disclosures firms have to consider more thoroughly what social information is financially material and must be included. Accordingly, the information disclosed via the SEC is likely more meaningful for the capital market due to its high level of specificity (Hope et al., 2016) and the provision of financially material information (Christensen et al., 2017). Acknowledging the risk function of non-financial information, SEC disclosures are an indication that the social information disclosed reveals risks that might have been previously unknown or unexpected. This argument is in line with the increased idiosyncratic risk that first-time SEC reporters experience.

### 5.3 Additional Analyses and Robustness Tests

We apply Propensity Score Matching to alleviate potential concerns about functional form misspecification (Shipman, Swanquist, & Whited, 2017). We also carry out Propensity Score Matching separately for each of the three social issues. For each social issue, we match one non-reporting firm to one first-time reporting firm in the same year of the first-time reporting based on the following estimation:

$$SEC_[Topic]_First = \omega_0 + \omega_1 Size + \omega_2 ROA + \omega_3 SDROA + \omega_4 Loss + \omega_5 MTB + \omega_6 AnCov + \omega_7 ESGScore + \omega_8 FinTran + \omega_9 FinTraInt + Industry\text{-}controls + Year\text{-}controls + \varepsilon_4 \quad (4)$$

We apply Propensity Score Matching without replacement based on one-to-one matches with a caliper distance of 0.03. Table 6 summarizes results for the Propensity Score Matched samples with idiosyncratic risk as the dependent variable (results for downside idiosyncratic risk look similar). As before, we find a consistent, positive, and highly significant ($p < 0.01$) coefficient for SEC_[Topic]_First across all models. Results for other variables are not consistent, but mostly insignificant. We also find further significant and positive coefficients of other variables of interest only for human capital: SUS_[Topic]_First (0.1823, $p < 0.01$) and SEC_[Topic]_Cont (0.1792, $p < 0.05$; and for the SEC channel only: 0.1935, $p < 0.01$).

[Insert Table 6 about here.]

Furthermore, we carry out a range of robustness tests to ensure the stability of our results regarding a number of research design choices. More specifically, we find qualitatively similar results when we apply a fixed effects model and apply additional or alternative control variables.

## 6. CONCLUSION

This study analyses the relation between first-time disclosures and idiosyncratic risk via three disclosure channels: SEC filings, sustainability reports, and financial reports. We argue that: (1) SEC disclosures are, on average, perceived as relevant and material information for (potential) investors, while the same does not necessarily hold true for other channels; and (2) sustainability-related disclosures are understood by capital markets mainly from a risk perspective. Accordingly, we hypothesized that first-time disclosures via SEC are related to increased uncertainty, measured as idiosyncratic risk.

The results support our main hypothesis by showing a consistent positive coefficient for first-time SEC disclosure for three distinct social disclosure issues and the aggregated social disclosure measure. The results hold for a range of different analyses and robustness tests and are supported by results for Propensity Score Matched samples. We find inconsistent and largely insignificant results for first-time disclosures via the other two disclosure channels, meaning sustainability and financial reports, whereas continued disclosure via these channels – most consistently via financial reports – is related to lower idiosyncratic risk. Overall, the results are in line with our argument that first-time disclosures via the SEC are of an unexpected and 'risky' nature. Our study thereby challenges the notion that sustainability disclosure impacts are independent of their disclosure channels. Our findings also carry direct policy relevance: as initially voluntary sustainability disclosures migrate into regulated channels – the trajectory attempted by the SEC's 2024 climate disclosure rules – first-time disclosure via a regulated, financially material channel can increase rather than reduce investor uncertainty.

The study's contribution to the literature is threefold. First, we extend the extant literature on disclosure channels (Blankespoor et al., 2020), risk disclosures (Campbell et al., 2014; Kravet & Muslu, 2013) and the use of sustainability information by investors (Christophe et al., 2024; Hawn, Chatterji, & Mitchell, 2018; Lu et al., 2021) by showing that investors' perception of social disclosures depends on the disclosure channel. Therefore, our results are in line with Christensen et al. (2017) by emphasizing that disclosures via regulatory filings are especially relevant for investors. Second, we add to the literature on the effects of disclosure regulation (Chen et al., 2018; Christensen et al., 2017; Grewal et al., 2019). Instead of focusing on a regulatory change, we simultaneously consider information disclosed via three different disclosure channels, and we add to the literature by revealing that these channels are differently linked to idiosyncratic firm risk. That means investors interpret the social information shared in the context of the disclosure channel used by the firm. Third, we add to the literature focusing on information economics (Brown et al., 2009; Diamond & Verrecchia, 1991; Kothari et al., 2009). Our study provides evidence that new sustainability information reported via SEC disclosures qualifies as unexpected information, which increases idiosyncratic risk.

*Declaration of generative AI and AI-assisted technologies in the manuscript preparation process:* During the preparation of this work the authors used Claude in order to review the manuscript. After using this tool/service, the authors reviewed and edited the content as needed and take full responsibility for the content of the published article.

## References

Aksoy, L., Buoye, A. J., Fors, M., Keiningham, T. L., & Rosengren, S. (2022). Environmental, Social and Governance (ESG) metrics do not serve services customers: a missing link between sustainability metrics and customer perceptions of social innovation. *Journal of Service Management*, *33*(4/5), 565–577. doi:10.1108/JOSM-11-2021-0428

Albuquerque, R., Koskinen, Y., & Zhang, C. (2019). Corporate Social Responsibility and Firm Risk: Theory and Empirical Evidence. *Management Science*, *65*(10), 4451–4469. doi:10.1287/mnsc.2018.3043

Aluchna, M., Roszkowska-Menkes, M., Kamiński, B., & Bosek-Rak, D. (2022). Do institutional investors encourage firm to social disclosure? The stakeholder salience perspective. *Journal of Business Research*, *142*, 674–682. doi:10.1016/j.jbusres.2021.12.064

Amel-Zadeh, A., & Serafeim, G. (2018). Why and How Investors Use ESG Information: Evidence from a Global Survey. *Financial Analysts Journal*, *74*(3), 87–103. doi:10.2469/faj.v74.n3.2

Arimond, A., Borth, D., Garcia Vega, S., Harjoto, M., Hoepner, A. G. F., Klawunn, M., & Weisheit, S. (2020). Neural Networks and Value at Risk. *SSRN Electronic Journal.* doi:10.2139/ssrn.3591996

Barth, M. E., Cahan, S. F., Chen, L., & Venter, E. R. (2017). The economic consequences associated with integrated report quality: Capital market and real effects. *Accounting, Organizations and Society*, *62*, 43–64. doi:10.1016/j.aos.2017.08.005

Beatty, A., Cheng, L., & Zhang, H. (2019). Are Risk Factor Disclosures Still Relevant? Evidence from Market Reactions to Risk Factor Disclosures Before and After the Financial Crisis. *Contemporary Accounting Research*, *36*(2), 805–838. doi:10.1111/1911-3846.12444

Becchetti, L., Ciciretti, R., & Hasan, I. (2015). Corporate social responsibility, stakeholder risk, and idiosyncratic volatility. *Journal of Corporate Finance*, *35*, 297–309. doi:10.1016/j.jcorpfin.2015.09.007

Benlemlih, M., & Girerd-Potin, I. (2017). Corporate social responsibility and firm financial risk reduction: On the moderating role of the legal environment. *Journal of Business Finance & Accounting*, *44*(7-8), 1137–1166. doi:10.1111/jbfa.12251

Benlemlih, M., Shaukat, A., Qiu, Y., & Trojanowski, G. (2018). Environmental and Social Disclosures and Firm Risk. *Journal of Business Ethics*, *152*(3), 613–626. doi:10.1007/s10551-016-3285-5

Berkman, H., Jona, J., & Soderstrom, N. S. (2024). Firm-Specific Climate Risk and Market Valuation. *Accounting, Organizations and Society*, *112*. doi:10.1016/j.aos.2024.101547

Bingler, J. A., Kraus, M., Leippold, M., & Webersinke, N. (2022). Cheap talk and cherry-picking: What ClimateBert has to say on corporate climate risk disclosures. *Finance Research Letters*, *47*, 102776. doi:10.1016/j.frl.2022.102776

Blacconiere, W. G., & Patten, D. M. (1994). Environmental disclosures, regulatory costs, and changes in firm value. *Journal of Accounting and Economics*, *18*(3), 357–377. doi:10.1016/0165-4101(94)90026-4

Blankespoor, E., deHaan, E., & Marinovic, I. (2020). Disclosure processing costs, investors' information choice, and equity market outcomes: A review. *Journal of Accounting and Economics*, *70*(2), 101344. doi:10.1016/j.jacceco.2020.101344

Bochkay, K., Choi, S., & Hales, J. (2022). When companies choose their reporting standards: Evidence on SASB adoption and associated outcomes. *SSRN Electronic Journal*. doi:10.2139/ssrn.4167391

Brown, S., Hillegeist, S. A., & Lo, K. (2009). The effect of earnings surprises on information asymmetry. *Journal of Accounting and Economics*, *47*(3), 208–225. doi:10.1016/j.jacceco.2008.12.002

Campbell, J. L., Chen, H., Dhaliwal, D. S., Lu, H., & Steele, L. B. (2014). The information content of mandatory risk factor disclosures in corporate filings. *Review of Accounting Studies*, *19*(1), 396–455. doi:10.1007/s11142-013-9258-3

Cazier, R. A., McMullin, J. L., & Treu, J. S. (2021). Are Lengthy and Boilerplate Risk Factor Disclosures Inadequate? An Examination of Judicial and Regulatory Assessments of Risk Factor Language. *The Accounting Review*, *96*(4), 131–155. doi:10.2308/TAR-2018-0657

Chen, M. A., Wu, Q., & Yang, B. (2019). How Valuable Is FinTech Innovation? *The Review of Financial Studies*, *32*(5), 2062–2106. doi:10.1093/rfs/hhy130

Chen, Y.-C., Hung, M., & Wang, Y. (2018). The effect of mandatory CSR disclosure on firm profitability and social externalities: Evidence from China. *Journal of Accounting and Economics*, *65*(1), 169–190. doi:10.1016/j.jacceco.2017.11.009

Chen, Z., & Xie, G. (2022). ESG disclosure and financial performance: Moderating role of ESG investors. *International Review of Financial Analysis*, *83*, 102291. doi:10.1016/j.irfa.2022.102291

Cho, S. Y., Lee, C., & Pfeiffer, R. J. (2013). Corporate social responsibility performance and information asymmetry. *Journal of Accounting and Public Policy*, *32*(1), 71–83. doi:10.1016/j.jaccpubpol.2012.10.005

Christensen, H. B., Floyd, E., Liu, L. Y., & Maffett, M. (2017). The real effects of mandated information on social responsibility in financial reports: Evidence from mine-safety records. *Journal of Accounting and Economics*, *64*(2), 284–304. doi:10.1016/j.jacceco.2017.08.001

Christensen, H. B., Hail, L., & Leuz, C. (2021). Mandatory CSR and sustainability reporting: economic analysis and literature review. *Review of Accounting Studies*, *26*(3), 1176–1248. doi:10.1007/s11142-021-09609-5

Christophe, S. E., Hsieh, J., & Lee, H. (2024). Reputation and recency: How do aggressive short sellers assess ESG-Related Information? *Journal of Business Research*, *180*, 114718. doi:10.1016/j.jbusres.2024.114718

Crowley, R., Huang, W., & Lu, H. (2024). Discretionary Dissemination on Twitter. *Contemporary Accounting Research*, *41*(4), 2454–2487. doi:10.1111/1911-3846.12986

Del Rossi, A. F., & Viscusi, W. K. (2010). The Changing Landscape of Blockbuster Punitive Damages Awards. *American Law and Economics Review*, *12*(1), 116–161. doi:10.1093/aler/ahp018

Dhaliwal, D. S., Li, O. Z., Tsang, A., & Yang, Y. G. (2011). Voluntary Nonfinancial Disclosure and the Cost of Equity Capital: The Initiation of Corporate Social Responsibility Reporting. *The Accounting Review*, *86*(1), 59–100. doi:10.2308/accr.00000005

Dhaliwal, D. S., Radhakrishnan, S., Tsang, A., & Yang, Y. G. (2012). Nonfinancial Disclosure and Analyst Forecast Accuracy: International Evidence on Corporate Social

Responsibility Disclosure. *The Accounting Review*, *87*(3), 723–759. doi:10.2308/accr-10218

Diamond, D. W., & Verrecchia, R. E. (1991). Disclosure, Liquidity, and the Cost of Capital. *The Journal of Finance*, *46*(4), 1325–1359. doi:10.1111/j.1540-6261.1991.tb04620.x

Elshandidy, T., Fraser, I., & Hussainey, K. (2013). Aggregated, voluntary, and mandatory risk disclosure incentives: Evidence from UK FTSE all-share companies. *International Review of Financial Analysis*, *30*, 320–333. doi:10.1016/j.irfa.2013.07.010

Fafaliou, I., Giaka, M., Konstantios, D., & Polemis, M. (2022). Firms' ESG reputational risk and market longevity: A firm-level analysis for the United States. *Journal of Business Research*, *149*, 161–177. doi:10.1016/j.jbusres.2022.05.010

Fama, E. F., & French, K. R. (1993). Common risk factors in the returns on stocks and bonds. *Journal of Financial Economics*, *33*(1), 3–56. doi:10.1016/0304-405X(93)90023-5

Fama, E. F., & French, K. R. (1996). Multifactor Explanations of Asset Pricing Anomalies. *The Journal of Finance*, *51*(1), 55–84. doi:10.1111/j.1540-6261.1996.tb05202.x

Farag, H., Meng, Q., & Mallin, C. (2015). The social, environmental and ethical performance of Chinese companies: Evidence from the Shanghai Stock Exchange. *International Review of Financial Analysis*, *42*, 53–63. doi:10.1016/j.irfa.2014.12.002

Freedman, M., & Patten, D. M. (2004). Evidence on the pernicious effect of financial report environmental disclosure. *Accounting Forum*, *28*(1), 27–41. doi:10.1016/j.accfor.2004.04.006

Fu, F. (2009). Idiosyncratic risk and the cross-section of expected stock returns. *Journal of Financial Economics*, *91*(1), 24–37. doi:10.1016/j.jfineco.2008.02.003

Gao, Y., Hoepner, A. G., Prokopczuk, M., Rouxelin, F., & Wuersig, C. (2025). Responsible investing: Upside potential and downside protection? *International Review of Financial Analysis*, *97*, 103754. doi:10.1016/j.irfa.2024.103754

Garber, S. (1998). Product liability, punitive damages, business decisions and economic outcomes. *Wisconsin Law Review*, 237–296. Retrieved from https://heinonline.org/HOL/P?h=hein.journals/wlr1998&i=257

Gelles, D. (2016, January 23). S.E.C. is criticized for lax enforcement of climate risk disclosure. *The New York Times*. Retrieved from https://www.nytimes.com/2016/01/24/business/energy-environment/sec-is-criticized-for-lax-enforcement-of-climate-risk-disclosure.html

Gillan, S. L., Koch, A., & Starks, L. T. (2021). Firms and social responsibility: A review of ESG and CSR research in corporate finance. *Journal of Corporate Finance*, *66*, 101889. doi:10.1016/j.jcorpfin.2021.101889

Göttsche, M., Griffin, P., Habermann, F., Schiemann, F., & Spandel, T. (2025). A Double-Edged Sword: Materiality Classifications of Sustainability Topics. *Review of Accounting Studies*, *30*(4), 3596–3639. doi:10.1007/s11142-025-09908-1

Grewal, J., Hauptmann, C., & Serafeim, G. (2021). Material Sustainability Information and Stock Price Informativeness. *Journal of Business Ethics*, *171*, 513–544. doi:10.1007/s10551-020-04451-2

Grewal, J., Riedl, E. J., & Serafeim, G. (2019). Market Reaction to Mandatory Nonfinancial Disclosure. *Management Science*, *65*(7), 3061–3084. doi:10.1287/mnsc.2018.3099

Gu, S., Kelly, B., & Xiu, D. (2020). Empirical Asset Pricing via Machine Learning. *The Review of Financial Studies*, *33*(5), 2223–2273. doi:10.1093/rfs/hhaa009

Guidry, R. P., & Patten, D. M. (2010). Market reactions to the first-time issuance of corporate sustainability reports. *Sustainability Accounting, Management and Policy Journal*, *1*(1), 33–50. doi:10.1108/20408021011059214

Hawn, O., Chatterji, A. K., & Mitchell, W. (2018). Do investors actually value sustainability? New evidence from investor reactions to the Dow Jones Sustainability Index (DJSI). *Strategic Management Journal*, *39*(4), 949–976. doi:10.1002/smj.2752

He, F., Ding, C., Yue, W., & Liu, G. (2023). ESG performance and corporate risk-taking: Evidence from China. *International Review of Financial Analysis*, *87*, 102550. doi:10.1016/j.irfa.2023.102550

He, F., Qin, S., Liu, Y., & Wu, J. (2022). CSR and idiosyncratic risk: Evidence from ESG information disclosure. *Finance Research Letters*, *49*, 102936. doi:10.1016/j.frl.2022.102936

Hoepner, A. G. F., Majoch, A. A. A., & Zhou, X. Y. (2021). Does an Asset Owner's Institutional Setting Influence Its Decision to Sign the Principles for Responsible Investment? *Journal of Business Ethics*, *168*(2), 389–414. doi:10.1007/s10551-019-04191-y

Hope, O.-K., Hu, D., & Lu, H. (2016). The benefits of specific risk-factor disclosures. *Review of Accounting Studies*, *21*(4), 1005–1045. doi:10.1007/s11142-016-9371-1

Hummel, K., & Schlick, C. (2016). The relationship between sustainability performance and sustainability disclosure – Reconciling voluntary disclosure theory and legitimacy theory. *Journal of Accounting and Public Policy*, *35*(5), 455–476. doi:10.1016/j.jaccpubpol.2016.06.001

Ibrahim, A. E. A., Hussainey, K., Nawaz, T., Ntim, C., & Elamer, A. (2022). A systematic literature review on risk disclosure research: State-of-the-art and future research agenda. *International Review of Financial Analysis*, *82*, 102217. doi:10.1016/j.irfa.2022.102217

Ignatov, K. (2023). When ESG talks: ESG tone of 10-K reports and its significance to stock markets. *International Review of Financial Analysis*, *89*, 102745. doi:10.1016/j.irfa.2023.102745

Ilhan, E., Krueger, P., Sautner, Z., & Starks, L. T. (2023). Climate Risk Disclosure and Institutional Investors. *The Review of Financial Studies*, *36*(7), 2617–2650. doi:10.1093/rfs/hhad002

Khan, M., Serafeim, G., & Yoon, A. (2016). Corporate Sustainability: First Evidence on Materiality. *The Accounting Review*, *91*(6), 1697–1724. doi:10.2308/accr-51383

Kim, S., Lee, G., & Kang, H.-G. (2021). Risk management and corporate social responsibility. *Strategic Management Journal*, *42*(1), 202–230. doi:10.1002/smj.3224

Kothari, S. P., Li, X., & Short, J. E. (2009). The Effect of Disclosures by Management, Analysts, and Business Press on Cost of Capital, Return Volatility, and Analyst Forecasts: A Study Using Content Analysis. *The Accounting Review*, *84*(5), 1639–1670. doi:10.2308/accr.2009.84.5.1639

Kravet, T., & Muslu, V. (2013). Textual risk disclosures and investors' risk perceptions. *Review of Accounting Studies*, *18*(4), 1088–1122. doi:10.1007/s11142-013-9228-9

Krueger, P., Sautner, Z., & Starks, L. T. (2020). The Importance of Climate Risks for Institutional Investors. *The Review of Financial Studies*, *33*(3), 1067–1111. doi:10.1093/rfs/hhz137

Lee, M. T., & Raschke, R. L. (2023). Stakeholder legitimacy in firm greening and financial performance: What about greenwashing temptations? *Journal of Business Research*, *155*, 113393. doi:10.1016/j.jbusres.2022.113393

Lin, B., Wang, X., Wu, J., & Wu, Y. (2026). Does subindustry disclosure policy affect auditor information supply? Evidence from China. *International Review of Financial Analysis*, *109*, 104776. doi:10.1016/j.irfa.2025.104776

Lin, Y., Shen, R., Wang, J., & Julia Yu, Y. (2024). Global Evolution of Environmental and Social Disclosure in Annual Reports. *Journal of Accounting Research*, *62*(5), 1941–1988. doi:10.1111/1475-679X.12575

Lu, H., Oh, W.-Y., Kleffner, A., & Chang, Y. K. (2021). How do investors value corporate social responsibility? Market valuation and the firm specific contexts. *Journal of Business Research*, *125*, 14–25. doi:10.1016/j.jbusres.2020.11.063

Matsumura, E. M., Prakash, R., & Vera-Muñoz, S. C. (2014). Firm-Value Effects of Carbon Emissions and Carbon Disclosures. *The Accounting Review*, *89*(2), 695–724. doi:10.2308/accr-50629

Matsumura, E. M., Prakash, R., & Vera-Muñoz, S. C. (2024). Climate-Risk Materiality and Firm Risk. *Review of Accounting Studies*, *29*(1), 33–74. doi:10.1007/s11142-022-09718-9

Melloni, G., Caglio, A., & Perego, P. (2017). Saying more with less? Disclosure conciseness, completeness and balance in Integrated Reports. *Journal of Accounting and Public Policy*, *36*(3), 220–238. doi:10.1016/j.jaccpubpol.2017.03.001

Moretti, G., Terzani, S., & De Novellis, G. (2026). Does Politics Influence Environmental, Social, and Governance Disclosure? Empirical Evidence From US Listed Firms. *Business Strategy and the Environment*, *35*(3), 4148–4170. doi:10.1002/bse.70373

Plumlee, M., Brown, D., Hayes, R. M., & Marshall, R. S. (2015). Voluntary environmental disclosure quality and firm value: Further evidence. *Journal of Accounting and Public Policy*, *34*(4), 336–361. doi:10.1016/j.jaccpubpol.2015.04.004

Rogers, J. L., Skinner, D. J., & van Buskirk, A. (2009). Earnings guidance and market uncertainty. *Journal of Accounting and Economics*, *48*(1), 90–109. doi:10.1016/j.jacceco.2009.07.001

Sassen, R., Hinze, A.-K., & Hardeck, I. (2016). Impact of ESG factors on firm risk in Europe. *Journal of Business Economics*, *86*(8), 867–904. doi:10.1007/s11573-016-0819-3

Schiemann, F., & Sakhel, A. (2019). Carbon Disclosure, Contextual Factors, and Information Asymmetry: The Case of Physical Risk Reporting. *European Accounting Review*, *28*(4), 791–818. doi:10.1080/09638180.2018.1534600

Schiemann, F., & Tietmeyer, R. (2022). ESG Controversies, ESG Disclosure and Analyst Forecast Accuracy. *International Review of Financial Analysis*, *84*, 102373. doi:10.1016/j.irfa.2022.102373

Schroeder, D., & Ormazabal, G. (2024). *E&S Disclosure Regulation and Political Ideology. SSRN Electronic Journal. doi:10.2139/ssrn.4986621*

Serafeim, G., & Yoon, A. (2022). Which Corporate ESG News Does the Market React To? Financial Analysts Journal, *78*(1), 59–78. doi.org/10.1080/0015198X.2021.1973879

Shipman, J. E., Swanquist, Q. T., & Whited, R. L. (2017). Propensity Score Matching in Accounting Research. *The Accounting Review*, *92*(1), 213–244. doi:10.2308/accr-51449

Thng, T. (2019). Do VC-backed IPOs manage tone? *The European Journal of Finance*, *25*(17), 1655–1682. doi:10.1080/1351847X.2018.1561482

Unerman, J., Bebbington, J., & O'Dwyer, B. (2007). Introduction to Sustainability Accounting and Accountability. In Jeffrey Unerman, Jan Bebbington, & Brendan O’Dwyer (Eds.), *Sustainability Accounting and Accountability* (pp. 1–16). London: Routledge.

Zhao, L., Yang, M. M., Wang, Z., & Michelson, G. (2023). Trends in the Dynamic Evolution of Corporate Social Responsibility and Leadership: A Literature Review and Bibliometric Analysis. *Journal of Business Ethics*, *182*(1), 135–157. doi:10.1007/s10551-022-05035-y

Zhou, J., & Lei, X. (2025). ESG rating uncertainty and corporate financial misconduct. *Journal of Business Research*, *189*, 115092. doi:10.1016/j.jbusres.2024.115092

Zuo, J., Zhang, W., Hu, M., Feng, X., & Zou, G. (2022). Employee relations and stock price crash risk: Evidence from employee lawsuits. *International Review of Financial Analysis*, *82*, 102188. doi:10.1016/j.irfa.2022.102188

## Figures

*Figure 1: Conceptual diagram on focus on financial materiality & regulatory enforcement regarding social disclosure of three disclosure channels*

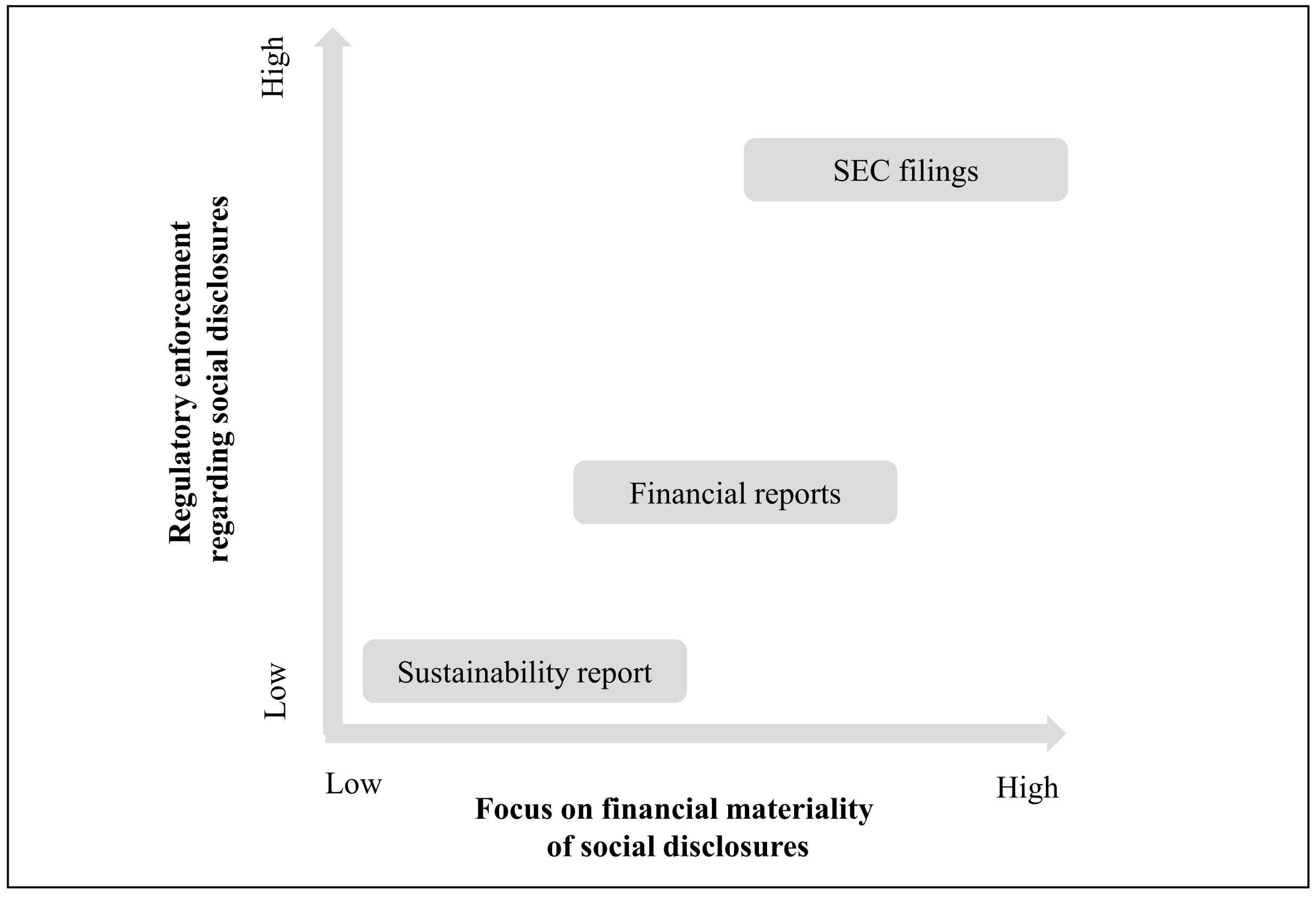

## Tables

*Table 1: Overview variable definitions*

| Variable | Name | Definition |
|---|---|---|
| Id_risk | Idiosyncratic risk | Squared residual of three factor model |
| neg_idrisk | Downside idiosyncratic risk | Squared residual of three factor model for negative market returns |
| Size | Firm size | Natural logarithm of total assets in USD |
| ROA | Return on assets | Pre-tax income divided by total assets |
| SDROA | St. dev of return on asset | Standard deviation of return on assets |
| Loss | Loss dummy | 1 if pre-tax income is negative; 0 otherwise |
| MTB | Market-to-book ratio | Market value of equity divided by book value of common shareholders' equity |
| AnCov | Analyst coverage | Number of analysts covering the respective company |
| ESGScore | ESG Score | Industry adjusted MSCI ESG score |
| SP500 | S&P 500 dummy | 1 if company is in S&P 500 in the respective year; 0 otherwise |
| FinTran | Financial Transparency | Ratio of items reported to items necessary to calculate scaled accruals |
| FinTraInt | Scaled accruals | Change in current assets minus change in current liabilities minus change in cash plus change in current portion of long term debt minus depreciations plus change in payment of income tax – divided by total assets of previous year; variable is 0 if not all of the necessary information is available |
| SEC_First<br>SUS_First<br>FIN_First | First disclosure via respective channel (SEC, SUS, FIN) | 1 if the company reports about at least one social topic via the respective channel (i.e., SEC, SUS, FIN) about which the company did not report in the previous year; 0 otherwise |
| SEC_Cont<br>SUS_Cont<br>FIN_Cont | Continued disclosure via respective channel (SEC, SUS, FIN) | 1 if the company reports about at least one social topic via the respective channel (i.e., SEC, SUS, FIN) about which the company did also report in the previous year; 0 otherwise |
| SEC_Topic_First<br>SUS_Topic_First<br>FIN_Topic_First | Topic-specific first disclosure via respective channel (SEC, SUS, FIN) | 1 if the company reports about the specific topic (i.e., Human Capital, Product Liabilities, Stakeholder Engagement) via the respective channel (i.e., SEC, SUS, FIN) and the company did not report about this topic in the previous year; 0 otherwise |
| SEC_Topic_Cont<br>SUS_Topic_Cont<br>FIN_Topic_Cont | Topic-specific continued disclosure via respective channel (SEC, SUS, FIN) | 1 if the company reports about the specific topic (i.e., Human Capital, Product Liabilities, Stakeholder Engagement) via the respective channel (i.e., SEC, SUS, FIN) and the company also reported about this topic in the previous year; 0 otherwise |

*Table 2: Descriptive Statistics*

**Panel A: Continuous variables**

| Variable name | Mean | Sd | p1 | p5 | p25 | p50 | p75 | p95 | p99 |
|---|---|---|---|---|---|---|---|---|---|
| Id_risk | 0.014 | 0.005 | 0.006 | 0.007 | 0.010 | 0.012 | 0.016 | 0.025 | 0.034 |
| Size | 15.685 | 1.364 | 12.879 | 13.638 | 14.679 | 15.550 | 16.615 | 18.077 | 19.002 |
| ROA | 0.095 | 0.082 | -0.163 | -0.015 | 0.045 | 0.088 | 0.138 | 0.240 | 0.344 |
| SDROA | 0.041 | 0.043 | 0.003 | 0.005 | 0.014 | 0.027 | 0.048 | 0.126 | 0.231 |
| Loss | 0.067 | 0.250 | 0 | 0 | 0 | 0 | 0 | 1 | 1 |
| MTB | 4.049 | 4.049 | 0.788 | 1.150 | 1.877 | 2.859 | 4.478 | 10.704 | 27.118 |
| AnCov | 2.711 | 1 | 1.099 | 1.609 | 2.398 | 2.833 | 3.135 | 3.526 | 3.738 |
| ESGScore | 4.406 | 2.143 | 0.000 | 1.300 | 2.860 | 4.200 | 5.900 | 8.200 | 9.800 |
| SP500 | 0.550 | 0.498 | 0 | 0 | 0 | 1 | 1 | 1 | 1 |
| FinTran | 0.923 | 0.090 | 0.667 | 0.750 | 0.833 | 1 | 1 | 1 | 1 |
| FinTraInt | -0.019 | 0.049 | -0.185 | -0.111 | -0.039 | 0.000 | 0.000 | 0.035 | 0.141 |

**Panel B: Frequencies of disclosure variables**

| | [Channel]_First | [Channel]_[Topic]_First | | |
|---|---|---|---|---|
| Channels | Overall | Human | Product | Stakeholder |
| SEC | 0.065 | 0.030 | 0.031 | 0.040 |
| SUS | 0.059 | 0.050 | 0.051 | 0.051 |
| FIN | 0.030 | 0.023 | 0.025 | 0.024 |

*Notes:*
*The table provides descriptive statistics of dependent and independent variables for the basic sample of 2,982 firm-year observations. Variable descriptions are provided in Table 1. Panel A reports descriptive statistics for continuous variables as indicated in column headings. "Mean" reports the average value, "Sd" reports the standard deviation, and "p1" to "p99" report respective percentile values.*
*Panel B provides frequencies for first-time disclosure variables. Thereby, each line focuses on one disclosure channel (i.e. SEC for SEC filings, SUS for sustainability reports, FIN for financial reports). Columns are dedicated to the disclosure topic, where Overall indicates the overall disclosure on social topics, Human indicates disclosures on human capital, Product indicates disclosure about product liabilities, and Stakeholder indicates disclosure about stakeholder engagement.*

*Table 3: Regression model with idiosyncratic risk as the dependent variable for the three disclosure channels individually and combined*

| | **(1) Basic Model** | | **(2) SEC** | | **(3) Sustainabiliy Report** | | **(4) Annual Report** | |
|---|---|---|---|---|---|---|---|---|
| | Coeff. | Std. Error | Coeff. | Std. Error | Coeff. | Std. Error | Coeff. | Std. Error |
| SEC_First | 0.0531 | (0.0136)*** | 0.0542 | (0.0138)*** | | | | |
| SUS_First | 0.0044 | (0.0193) | | | 0.0070 | (0.0206) | | |
| FIN_First | -0.0226 | (0.0242) | | | | | -0.0188 | (0.0239) |
| SEC_Cont | -0.0387 | (0.0342) | -0.0318 | (0.0339) | | | | |
| SUS_Cont | -0.0338 | (0.0174)* | | | -0.0422 | (0.0171)** | | |
| FIN_Cont | -0.0542 | (0.0309)* | | | | | -0.0592 | (0.0299)** |
| Size | -0.1620 | (0.0178)*** | -0.1765 | (0.0187)*** | -0.1694 | (0.0176)*** | -0.1640 | (0.0179)*** |
| ROA | -1.5411 | (0.2028)*** | -1.5849 | (0.206)*** | -1.5708 | (0.2037)*** | -1.5442 | (0.2003)*** |
| SDROA | 3.0477 | (0.3857)*** | 3.0447 | (0.3816)*** | 3.0867 | (0.3775)*** | 3.0991 | (0.3795)*** |
| Loss | 0.2604 | (0.0328)*** | 0.2553 | (0.0322)*** | 0.2598 | (0.0331)*** | 0.2612 | (0.0333)*** |
| MTB | -0.0019 | (0.0035) | -0.0026 | (0.0036) | -0.0023 | (0.0037) | -0.0018 | (0.0036) |
| AnCov | 0.2351 | (0.0354)*** | 0.2382 | (0.0354)*** | 0.2353 | (0.0353)*** | 0.2350 | (0.0348)*** |
| ESGScore | -0.0051 | (0.006) | -0.0070 | (0.006) | -0.0063 | (0.0062) | -0.0064 | (0.0058) |
| SP500 | -0.0495 | (0.0394) | -0.0468 | (0.04) | -0.0439 | (0.0387) | -0.0502 | (0.0397) |
| FinTran | -0.0646 | (0.1112) | -0.0725 | (0.1092) | -0.0675 | (0.1089) | -0.0675 | (0.1112) |
| FinTraInt | 0.2227 | (0.1981) | 0.2546 | (0.1965) | 0.2662 | (0.1981) | 0.2470 | (0.1994) |
| Constant | 3.4180 | (0.2267)*** | 3.6331 | (0.2288)*** | 3.5142 | (0.2194)*** | 3.4305 | (0.2239)*** |
| Industry controls | YES | | YES | | YES | | YES | |
| Year controls | YES | | YES | | YES | | YES | |
| N | 2,982 | | 2,982 | | 2,982 | | 2,982 | |
| $R^2$ | 0.4226 | | 0.4206 | | 0.419 | | 0.4198 | |
| Adj. $R^2$ | 0.4175 | | 0.4163 | | 0.4146 | | 0.4155 | |
| F | 72.6889 | *** | 84.3848 | *** | 83.6505 | *** | 84.8276 | *** |

*Notes:*
*The table reports regression results for four models with idiosyncratic risk as the dependent variable. The Basic Model in column (1) estimates all three disclosure channels simultaneously, while columns (2) to (4) focus on each disclosure channel (i.e., SEC, Sustainability Report, Annual Report) separately.*
*The table reports coefficients. Firm and year clustered standard errors are shown in parentheses. Variable definitions are shown in Table 1. SEC_Cont, SUS_Cont and FIN_Cont are additional control variables, which capture any continued reporting via the respective channel. Accordingly, each variable takes a value of 1 if a company reports about at least one social topic on which it already reported in the previous year via the respective channel. Otherwise, the variable is 0.*
*** $p < 0.01$, ** $p < 0.05$, * $p < 0.1$

*Table 4: Regression model with idiosyncratic risk as the dependent variable for three social topics and three disclosure channels*

| | *Idiosyncratic risk* | | | *Downside idiosyncratic risk* | | |
|---|---|---|---|---|---|---|
| Variable | Human Capital | Product Liabilities | Stakeholder Engagement | Human Capital | Product Liabilities | Stakeholder Engagement |
| SEC_[Topic]_First | 0.2494 | 0.2409 | 0.0953 | 0.2392 | 0.2175 | 0.1414 |
| | (0.0254)*** | (0.0352)*** | (0.0435)** | (0.0487)*** | (0.0319)*** | (0.0497)*** |
| SUS_[Topic]_First | -0.0133 | -0.0257 | -0.0162 | 0.1684 | 0.0552 | 0.0063 |
| | (0.0271) | (0.0362) | (0.0263) | (0.1941) | (0.1141) | (0.0572) |
| FIN_[Topic]_First | -0.0345 | -0.0451 | -0.0456 | -0.323 | -0.1513 | -0.2494 |
| | (0.0199)* | (0.0232)* | (0.0264)* | (0.2083) | (0.1513) | (0.1274)* |
| SEC_[Topic]_Cont | 0.0568 | 0.0321 | -0.0329 | 0.0822 | 0.0349 | -0.021 |
| | (0.0307)* | (0.0251) | (0.0331) | (0.0228)*** | (0.0304) | (0.0302) |
| SUS_[Topic]_Cont | -0.0188 | -0.0111 | -0.0216 | -0.0084 | 0.0009 | -0.0106 |
| | (0.023) | (0.0191) | (0.024) | (0.0254) | (0.0218) | (0.0268) |
| FIN_[Topic]_Cont | -0.0496 | -0.0564 | -0.0554 | -0.0265 | -0.0332 | -0.0276 |
| | (0.028)* | (0.0269)** | (0.029)* | (0.026) | (0.0252) | (0.0267) |
| Size | -0.1641 | -0.1639 | -0.1625 | -0.153 | -0.1542 | -0.1526 |
| | (0.0171)*** | (0.0174)*** | (0.0169)*** | (0.0163)*** | (0.0175)*** | (0.0162)*** |
| ROA | -1.5235 | -1.5413 | -1.5606 | -1.3729 | -1.4025 | -1.4171 |
| | (0.2109)*** | (0.2085)*** | (0.2073)*** | (0.1965)*** | (0.1962)*** | (0.1884)*** |
| SDROA | 3.0217 | 3.0512 | 3.0664 | 3.0254 | 3.0354 | 3.0565 |
| | (0.401)*** | (0.3836)*** | (0.3957)*** | (0.5748)*** | (0.5571)*** | (0.5675)*** |
| Loss | 0.2666 | 0.2664 | 0.2583 | 0.2811 | 0.2789 | 0.2705 |
| | (0.0319)*** | (0.0323)*** | (0.0319)*** | (0.0461)*** | (0.0453)*** | (0.0464)*** |
| MTB | -0.0021 | -0.0023 | -0.0019 | -0.0037 | -0.004 | -0.0038 |
| | (0.0037) | (0.0037) | (0.0037) | (0.0035) | (0.0035) | (0.0035) |
| AnCov | 0.2375 | 0.2378 | 0.2373 | 0.211 | 0.2131 | 0.2117 |
| | (0.0349)*** | (0.035)*** | (0.0355)*** | (0.0352)*** | (0.0355)*** | (0.0356)*** |
| ESGScore | -0.0057 | -0.0065 | -0.006 | -0.0048 | -0.0058 | -0.0051 |
| | (0.0064) | (0.0062) | (0.0063) | (0.0079) | (0.0075) | (0.0077) |
| SP500 | -0.0448 | -0.048 | -0.0476 | -0.064 | -0.0682 | -0.0684 |
| | (0.0402) | (0.0406) | (0.04) | (0.037)* | (0.0373)* | (0.0374)* |
| FinTran | -0.0771 | -0.0738 | -0.0677 | -0.0964 | -0.0925 | -0.0913 |
| | (0.1081) | (0.1088) | (0.109) | (0.1293) | (0.1274) | (0.1299) |
| FinTraInt | 0.2039 | 0.2084 | 0.2375 | 0.0867 | 0.0854 | 0.1098 |
| | (0.2013) | (0.2049) | (0.2015) | (0.0795) | (0.0773) | (0.0863) |
| Constant | 3.3838 | 3.4032 | 3.423 | 3.2322 | 3.2872 | 3.3046 |
| | (0.2259)*** | (0.2256)*** | (0.2185)*** | (0.2203)*** | (0.2281)*** | (0.2227)*** |
| | | | | | | |
| Industry controls | YES | YES | YES | YES | YES | YES |
| Year controls | YES | YES | YES | YES | YES | YES |
| | | | | | | |
| N | 2,982 | 2,982 | 2,982 | 2,982 | 2,982 | 2,982 |
| $R^2$ | 0.4239 | 0.4244 | 0.4221 | 0.3784 | 0.3772 | 0.3766 |
| Adj. $R^2$ | 0.4189 | 0.4193 | 0.417 | 0.3730 | 0.3717 | 0.3711 |
| F | 72.2564 | 73.2282 | 72.0803 | 62.6869 | 62.712 | 62.7239 |

*Notes:*
*The table reports regression results for six models with idiosyncratic risk as the dependent variable for the three models on the left, and negative idiosyncratic risk as the dependent variable for the three models on the right. Each regression model focuses on firm-reporting about a specific social topic (i.e., Human Capital, Product Liabilities, Stakeholder Engagement) and estimates all three disclosure channels simultaneously.*
*The table reports coefficients. Firm and year clustered standard errors are shown in parentheses. Variable definitions are shown in Table 1. SEC_Topic_Cont, SUS_Topic_Cont and FIN_Topic_Cont are additional control variables, which capture any continued reporting via the respective channel and about the specific social topic. Accordingly, each variable takes a value of 1 if a company reports about the respective social topic in the current year and in the previous year via the respective channel. Otherwise, the variable is 0.*
**** $p < 0.01$, ** $p < 0.05$, * $p < 0.1$*

Table 5: Regression model focusing on first-time SEC disclosures with idiosyncratic risk as the dependent variable for three social topics

| | **Human Capital** | | **Product Liabilities** | | **Stakeholder Engagement** | |
|---|---|---|---|---|---|---|
| | Coeff. | Std. Error | Coeff. | Std. Error | Coeff. | Std. Error |
| SEC_[Topic]_First | 0.2393 | (0.041)*** | 0.2587 | (0.0197)*** | 0.1187 | (0.044)*** |
| SUS_[Topic]_First | -0.0336 | (0.027) | -0.0419 | (0.0309) | -0.0352 | (0.0301) |
| FIN_[Topic]_First | -0.0264 | (0.0136)* | -0.0427 | (0.0165)*** | -0.0420 | (0.0249)* |
| SEC_[Topic]_First x SUS_[Topic]_First | 0.5189 | (0.1105)*** | 0.3600 | (0.1497)** | 0.2559 | (0.0842)*** |
| SEC_[Topic]_First x FIN_[Topic]_First | -0.2741 | (0.3154) | -0.1784 | (0.4325) | -0.1270 | (0.2606) |
| SEC_[Topic]_Cont | 0.0568 | (0.0307)* | 0.0331 | (0.0249) | -0.0337 | (0.0332) |
| SUS_[Topic]_Cont | -0.0205 | (0.0217) | -0.0085 | (0.0183) | -0.0222 | (0.0236) |
| FIN_[Topic]_Cont | -0.0429 | (0.0281) | -0.0487 | (0.0262)* | -0.0452 | (0.0294) |
| SEC_[Topic]_First x SUS_[Topic]_Cont | 0.0677 | (0.2849) | -0.1201 | (0.1924) | -0.0653 | (0.0375)* |
| SEC_[Topic]_First x FIN_[Topic]_Cont | -0.2444 | (0.3199) | -0.2063 | (0.1166)* | -0.2483 | (0.1077)** |
| Size | -0.1638 | (0.0174)*** | -0.1639 | (0.0174)*** | -0.1624 | (0.0169)*** |
| ROA | -1.5326 | (0.2134)*** | -1.5501 | (0.2097)*** | -1.5679 | (0.2092)*** |
| SDROA | 2.9934 | (0.4209)*** | 3.0177 | (0.3963)*** | 3.0435 | (0.4016)*** |
| Loss | 0.2676 | (0.0319)*** | 0.2679 | (0.0312)*** | 0.2588 | (0.0315)*** |
| MTB | -0.0020 | (0.0038) | -0.0023 | (0.0038) | -0.0020 | (0.0037) |
| AnCov | 0.2363 | (0.0353)*** | 0.2379 | (0.0358)*** | 0.2368 | (0.0359)*** |
| ESGScore | -0.0058 | (0.0063) | -0.0067 | (0.0062) | -0.0060 | (0.0062) |
| SP500 | -0.0450 | (0.0401) | -0.0488 | (0.0406) | -0.0476 | (0.0402) |
| FinTran | -0.0763 | (0.1071) | -0.0715 | (0.1089) | -0.0739 | (0.1075) |
| FinTraInt | 0.2167 | (0.2001) | 0.2147 | (0.2024) | 0.2321 | (0.2104) |
| Constant | 3.3856 | (0.2269)*** | 3.4026 | (0.2239)*** | 3.4302 | (0.2191)*** |
| Industry controls | YES | | YES | | YES | |
| Year controls | YES | | YES | | YES | |
| N | 2,982 | | 2,982 | | 2,982 | |
| $R^2$ | 0.426 | | 0.4263 | | 0.424 | |
| Adj. $R^2$ | 0.4202 | | 0.4205 | | 0.4181 | |
| F | 63.6643 | *** | 64.4887 | *** | 63.4744 | *** |

*Notes:*
*The table reports regression results for three models with idiosyncratic risk as the dependent variable. Each regression model focuses on firm-reporting about a specific social topic (i.e., Human Capital, Product Liabilities, Stakeholder Engagement) and estimates all three disclosure channels simultaneously. Thereby, the main focus is on the interaction terms SEC_Topic_First * SUS_Topic_First and SEC_Topic_First * FIN_Topic_First.*
*The table reports coefficients. Firm and year clustered standard errors are shown in parentheses. Variable definitions are shown in Table 1. SEC_Topic_Cont, SUS_Topic_Cont and FIN_Topic_Cont are additional control variables, which capture any continued reporting via the respective channel and about the specific social topic. Accordingly, each variable takes a value of 1 if a company reports about the respective social topic in the current year and in the previous year via the respective channel. Otherwise, the variable is 0.*
**** $p < 0.01$, ** $p < 0.05$, * $p < 0.1$*

*Table 6: Regression model for propensity score matched samples for SEC-disclosures with idiosyncratic risk as the dependent variable and for three social topics*

| | *All channels* | | | *SEC channel only* | | |
|---|---|---|---|---|---|---|
| Variable | Human Capital | Product Liabilities | Stakeholder Engagement | Human Capital | Product Liabilities | Stakeholder Engagement |
| SEC_[Topic]_First | 0.3726 | 0.267 | 0.1488 | 0.3937 | 0.2622 | 0.1553 |
| | (0.0513)*** | (0.0477)*** | (0.0474)*** | (0.046)*** | (0.0438)*** | (0.0523)*** |
| SUS_[Topic]_First | 0.1823 | 0.0563 | 0.1185 | | | |
| | (0.0688)*** | (0.0643) | (0.0829) | | | |
| FIN_[Topic]_First | -0.0298 | 0.027 | -0.0332 | | | |
| | (0.0671) | (0.1333) | (0.0808) | | | |
| SEC_[Topic]_Cont | 0.1792 | 0.0693 | -0.0039 | 0.1935 | 0.0705 | -0.0044 |
| | (0.0751)** | (0.0555) | (0.0449) | (0.0737)*** | (0.0495) | (0.046) |
| SUS_[Topic]_Cont | -0.0472 | 0.0266 | -0.0223 | | | |
| | (0.0699) | (0.0642) | (0.0343) | | | |
| FIN_[Topic]_Cont | -0.0608 | -0.165 | -0.0733 | | | |
| | (0.0618) | (0.1052) | (0.077) | | | |
| Size | -0.1544 | -0.1242 | -0.1783 | -0.1699 | -0.1425 | -0.1958 |
| | (0.0387)*** | (0.0449)*** | (0.0295)*** | (0.0315)*** | (0.0319)*** | (0.0281)*** |
| ROA | -0.7879 | -0.6775 | -1.4763 | -0.8034 | -0.6929 | -1.5343 |
| | (0.4379)* | (0.4147) | (0.2497)*** | (0.4523)* | (0.404)* | (0.2208)*** |
| SDROA | 1.5173 | 2.6833 | 3.2272 | 1.5782 | 2.6855 | 3.2147 |
| | (0.6805)** | (0.528)*** | (0.4492)*** | (0.652)** | (0.5328)*** | (0.4899)*** |
| Loss | 0.3157 | 0.2632 | 0.0874 | 0.3094 | 0.2619 | 0.0803 |
| | (0.0872)*** | (0.1184)** | (0.0582) | (0.0937)*** | (0.1239)** | (0.0596) |
| MTB | 0.0029 | -0.0011 | -0.0061 | 0.002 | -0.0023 | -0.0068 |
| | (0.0108) | (0.0093) | (0.0026)** | (0.0105) | (0.009) | (0.0025)*** |
| AnCov | 0.2168 | 0.2258 | 0.3198 | 0.2298 | 0.2345 | 0.323 |
| | (0.0799)*** | (0.085)*** | (0.0365)*** | (0.0799)*** | (0.0821)*** | (0.0355)*** |
| ESGScore | 0.0014 | -0.0047 | -0.0044 | 0 | -0.0068 | -0.0057 |
| | (0.0147) | (0.0118) | (0.0081) | (0.015) | (0.0117) | (0.0083) |
| SP500 | -0.199 | -0.258 | -0.1046 | -0.2 | -0.2353 | -0.0902 |
| | (0.1114)* | (0.1025)** | (0.0594)* | (0.1133)* | (0.0994)** | (0.0565) |
| FinTran | -0.2346 | 0.1204 | -0.2787 | -0.2704 | 0.1265 | -0.2471 |
| | (0.3464) | (0.251) | (0.2525) | (0.3229) | (0.2485) | (0.2689) |
| FinTraInt | 0.4517 | 1.0236 | 0.2892 | 0.4581 | 1.0443 | 0.3612 |
| | (0.7546) | (0.3975)** | (0.2882) | (0.7536) | (0.3953)*** | (0.2971) |
| Constant | 3.6894 | 2.7169 | 3.7229 | 3.916 | 2.9728 | 3.9553 |
| | (0.7821)*** | (0.7864)*** | (0.5372)*** | (0.6612)*** | (0.6274)*** | (0.5344)*** |
| Industry controls | YES | YES | YES | YES | YES | YES |
| Year controls | YES | YES | YES | YES | YES | YES |
| N | 497 | 494 | 601 | 497 | 494 | 601 |
| $R^2$ | 0.4518 | 0.4801 | 0.5146 | 0.4468 | 0.4725 | 0.5104 |
| Adj. R2 | 0.4215 | 0.4512 | 0.4927 | 0.4211 | 0.4479 | 0.4918 |
| F | 17.3383 | 21.9677 | 28.5463 | 18.5076 | 24.864 | 32.8575 |

*Notes:*
*The table reports regression results for six models with idiosyncratic risk as the dependent variable based on propensity score matched samples. Each regression model focuses on firm-reporting about a specific social topic (i.e., Human Capital, Product Liabilities, Stakeholder Engagement). The three models on the left estimate all three disclosure channels simultaneously, while the three models on the right focus on the SEC disclosures. The table reports coefficients. Firm and year clustered standard errors are shown in parentheses. Variable definitions are shown in Table 1. SEC_Topic_Cont, SUS_Topic_Cont and FIN_Topic_Cont are additional control variables, which capture any continued reporting via the respective channel and about the specific social topic. Accordingly, each variable takes a value of 1 if a company reports about the respective social topic in the current year and in the previous year via the respective channel. Otherwise, the variable is 0.*
**** $p < 0.01$, ** $p < 0.05$, * $p < 0.1$*